\documentclass[%
 reprint,
superscriptaddress,
 amsmath,amssymb,
 aps,
]{revtex4-1}

\usepackage{makecell}
\usepackage{multirow}
\usepackage{rotating}
\renewcommand{\rotcell}[1]{\rotatebox[origin=c]{90}{\parbox{2.8cm}{\centering #1}}}

\usepackage{graphicx}
\usepackage{dcolumn}
\usepackage{bm}
\usepackage{hyperref}
\usepackage[mathlines]{lineno}
\usepackage{xcolor}
\usepackage{float}

\begin{document}
\title{Multilayer network science: theory, methods, and applications}

\author{Alberto Aleta}
\affiliation{Institute for Biocomputation and Physics of Complex Systems (BIFI), University of Zaragoza, 50018, Zaragoza, Spain}
\affiliation{Department of Theoretical Physics, University of Zaragoza, 50018, Zaragoza, Spain}
\author{Andreia Sofia Teixeira}
\affiliation{BRAN Lab, Network Science Institute, Northeastern University London, London E1W 1LP, United Kingdom}
\affiliation{LASIGE, Faculdade de Ciências, Universidade de Lisboa, Lisbon 1749-016, Portugal}
\author{Guilherme Ferraz de Arruda}
\affiliation{Institute of Physics Gleb Wataghin, University of Campinas, Campinas 13083-970, Brazil}
\author{Andrea Baronchelli}
\affiliation{City St George’s, University of London, London EC1V 0HB, United Kingdom}
\author{Alain Barrat}
\affiliation{Turing Center for Living Systems, Aix-Marseille University, Université de Toulon, CNRS, CPT, Marseille 13009, France}
\author{J{\'a}nos Kert{\'e}sz}
\affiliation{Department of Network and Data Science, Central European University, Vienna 1100, Austria}
\author{Albert D{\'i}az-Guilera}
\affiliation{Departament de Física de la Mat{\`e}ria Condensada, Universitat de Barcelona, Barcelona 08028, Spain}
\affiliation{Universitat de Barcelona Institute of Complex Systems (UBICS), Universitat de Barcelona, Barcelona 08028, Spain}
\author{Oriol Artime}
\affiliation{Departament de Física de la Mat{\`e}ria Condensada, Universitat de Barcelona, Barcelona 08028, Spain}
\affiliation{Universitat de Barcelona Institute of Complex Systems (UBICS), Universitat de Barcelona, Barcelona 08028, Spain}
\author{Michele Starnini}
\affiliation{Department of Engineering, Universitat Pompeu Fabra, Barcelona 08028, Spain}
\author{Giovanni Petri}
\affiliation{Network Science Institute, Northeastern University London, London E1W 1LP, United Kingdom}
\author{Márton Karsai}
\affiliation{Department of Network and Data Science, Central European University, Vienna 1100, Austria}
\affiliation{National Laboratory for Health Security, HUN-REN Rényi Institute of Mathematics, Budapest 1053, Hungary}
\author{Siddharth Patwardhan}
\affiliation{Kellogg School of Management, Northwestern University, Evanston, IL, 60208, United States}
\author{Kathryn Coronges}
\affiliation{Network Science Institute, Northeastern University, Boston, MA, 02115, United States}
\author{Ann McCranie}
\affiliation{The Irsay Institute, Indiana University Bloomington, Bloomington, IN, 47405-3700, United States}
\author{Alessandro Vespignani}
\affiliation{Laboratory for the Modeling of Biological and Socio-technical Systems, Northeastern University, Boston, MA, 02115, United States}
\affiliation{Institute for Scientific Interchange Foundation, Turin, 10126, Italy}
\author{Yamir Moreno}
\affiliation{Institute for Biocomputation and Physics of Complex Systems (BIFI), University of Zaragoza, 50018, Zaragoza, Spain}
\affiliation{Department of Theoretical Physics, University of Zaragoza, 50018, Zaragoza, Spain}
\author{Santo Fortunato}
\email{santo@iu.edu}
\affiliation{Center for Complex Networks and Systems Research, Luddy School of Informatics, Computing, and Engineering, Indiana University Bloomington, Bloomington, IN, 47408, United States}



\date{\today}

\begin{abstract}
Multilayer network science has emerged as a central framework for analysing interconnected and interdependent complex systems. Its relevance has grown substantially with the increasing availability of rich, heterogeneous data, which makes it possible to uncover and exploit the inherently multilayered organisation of many real-world networks. In this review, we summarise recent developments in the field. On the theoretical and methodological front, we outline core concepts and survey advances in community detection, dynamical processes, temporal networks, higher-order interactions, and machine-learning-based approaches. On the application side, we discuss progress across diverse domains, including interdependent infrastructures, spreading dynamics, computational social science, economic and financial systems, ecological and climate networks, science-of-science studies, network medicine, and network neuroscience. We conclude with a forward-looking perspective, emphasizing the need for standardised datasets and software, deeper integration of temporal and higher-order structures, and a transition toward genuinely predictive models of complex systems.

\end{abstract}

\maketitle

\section{Introduction}

During the last 25 years, network science has established itself as a powerful and seamlessly integrated research discipline~\cite{dorogovtsev03b,vespignani04,caldarelli07,barrat08,cohen10,newman10,estrada11b,estrada15,barabasi16,latora17,menczer20,bagrow24}. At variance with other, more niched disciplines, it has allowed us to enrich and to synergize with other branches of science: from sociology and economics to biology, chemistry, and physics. This has been possible because of the data revolution and increased computational power that took place in the twilight of the 20th century. Nowadays, applications of network science tools pervade all fields of science but, most importantly, the discipline has changed the way we think about interactions in natural and human-made complex systems. 

The momentum for the development of network science came with the discovery that seemingly diverse networks have universal properties that pervade many natural and artificial systems. Since 1998, ignited by the works of Watts and Strogatz and Barab\'asi and Albert \cite{Watts1998Jun,Barabasi1999Oct}, systems as diverse as electric power grids, food chains, brain networks, protein networks, transcriptional networks, and social networks have been cast into the mathematical language of networks. We are now familiar with jargon that includes centrality, homogeneous and heterogeneous systems, communities, structural and dynamic phase transitions, weak and strong links, etc.

Yet, as the field matured, it became increasingly clear that most real systems cannot be faithfully represented as individual networks. Many natural, social, and technological systems are instead composed of several types of interactions, giving rise to multilayer or interdependent structures~\cite{kivela14,boccaletti14,battiston14,bianconi18,Aleta2019Mar,artime22,DeDomenico_2023}. The conceptual roots of this idea trace back to the sociological notion of multiplexity, where individuals were known to maintain several types of social ties simultaneously within the same population~\cite{Wasserman1994Nov}. These early multi-relational studies anticipated many of the ideas later formalized in network science. Then, with the rise of large-scale digital data around 2010, empirical analyses began to demonstrate that distinct layers of interaction can exhibit markedly different structural and dynamical patterns, underscoring the need for explicit multilayer representations~\cite{Szell2010Aug}.

A major turning point came with the study of interdependent networks, where nodes in one layer depend on nodes in another layer for functionality~\cite{buldyrev_2010,Vespignani2010Apr}. This framework revealed that coupling multiple networks can fundamentally alter systemic behaviour, producing abrupt percolation transitions and cascading failures, phenomena not observed in isolated networks. Following these foundational insights, the field rapidly matured from focusing primarily on cascades and robustness to developing a comprehensive mathematical formalism and exploring a wide array of dynamic processes. Researchers worked to create a unified language and a robust mathematical framework to describe the diverse architectures of multilayer systems~\cite{kivela14,boccaletti14,battiston14}, enabling the extension of classical concepts of network science to the multilayer context, such as centrality measures, among others. This theoretical consolidation sparked extensive research on dynamics across layers, including diffusion, synchronisation, epidemic spreading, and evolutionary games, revealing that interlayer coupling can either enhance or inhibit dynamical processes depending on structural correlations.

In the following years, multilayer methodologies spread rapidly across disciplines. As we will discuss in the following sections, applications now span social, biological, and technological systems. Parallel advances in community detection and inference have made it possible to extract meaningful mesoscale organisation from high-dimensional, heterogeneous data. More recent developments integrate temporal, categorical, and hierarchical aspects within multilayer frameworks and connect them to higher-order network representations such as hypergraphs. Today, the multilayer paradigm stands as one of the most powerful generalizations in network science, bridging theoretical, computational, and empirical approaches to the study of interconnected complex systems.

As such, multilayer network science has become a mature field. In this paper, we provide a high-level overview of progress in this area. We divide the content into two broad parts. In the first, about theory and methods, after a brief introduction of the main concepts, we will review the main techniques to detect communities in multilayer networks, show how dynamic processes unfold on these peculiar networks, summarize progress in the analysis, modelling, and dynamics of temporal networks, explain how higher-order structures can be used to study multilayer networks and how the latter can be projected in vector spaces via neural embeddings. The second part presents the applications of multilayer network science to diverse domains and topics. These include: interdependent systems, spreading processes, computational social science, economic and financial systems, infrastructure and ecological networks, science of science, climate science, network medicine, and network neuroscience.
At the end, we highlight current challenges and promising future research directions.

\section{Theory and Methods}

\subsection{Fundamental Concepts}
\subsubsection{Definitions and Formalisms}
\label{sec:definit}

A \textit{network}, or \textit{graph}, is a mathematical structure used to model pairwise relations between objects. It consists of a set of \textit{nodes} (or \textit{vertices}) and a set of \textit{links} (or \textit{edges}) that connect pairs of nodes. These connections are encoded in the \textit{adjacency matrix} \textbf{A}, where the element $a_{ij}=1$ if there is a link between nodes $i$ and $j$, and $a_{ij}=0$ otherwise.

A \textit{multilayer network} generalizes this concept to represent systems that integrate multiple subsystems or types of connectivity. Formally, it is a set of networks, with each network representing a \textit{layer} of the system. The layers can represent different times, contexts, or types of interaction. Importantly, there can be links within those layers, representing \textit{intralayer} connections, but also links connecting nodes across layers, \textit{interlayer} connections.

A prominent example of a multilayer system is the coupling of a power grid and the internet. One layer represents the power grid, with nodes as power stations and substations, while links represent transmission lines, illustrating the distribution of electrical power. Meanwhile, the other layer represents the internet, with nodes as data centres, routing hubs, and servers, linked by data transmission pathways. This setup highlights the interdependency between layers: the internet needs power to function, while modern power grids rely on internet-based communication for their operation, something that can be encoded in the interlayer links. This multilayer network model is essential for studying cascading failures and the overall resilience of interconnected systems \cite{buldyrev_2010, DeDomenico_2023}.

A \textit{multiplex network} is a specific and widely used type of multilayer network where nodes represent the same entity on every layer, and interlayer links only connect a node to itself across different layers. Each layer then represents a different type of relationship among the nodes. For instance, a social network can be modelled as a multiplex network where layers represent different platforms or social contexts~\cite{kivela14,battiston14}.

To analyse multiplex networks, we need to generalize the adjacency matrix. Two primary formalisms are used. First, we have the tensor formalism, in which the separation of layers can be maintained explicitly. A multiplex network can then be represented by an \textit{adjacency tensor} $\mathcal{M}$ of order 4 with elements $M_{ij}^{\alpha\beta}$. Here, $M_{ij}^{\alpha\alpha}$ represents the weight of the intralayer link between nodes $i$ and $j$ in layer $\alpha$, while $M_{ij}^{\alpha\beta}$ with $\alpha\neq\beta$ represents the interlayer connections. This is a more general and flexible representation, though it requires methods from tensor algebra for analysis~\cite{DeDomenico2013}.

Second, we have the formalism based on the \textit{supra-adjacency matrix}, which represents the entire multilayer system as a single, larger network. For a multiplex network with $N$ nodes and $L$ layers, the supra-adjacency matrix $\mathcal{A}$ is a $NL\times NL$ block matrix. The diagonal blocks are the adjacency matrices of each layer $\alpha$, $\text{\textbf{A}}^\alpha$, representing intralayer connections. The off-diagonal blocks encode the interlayer connections, which can be represented by the identity matrix \textbf{I} in the $\alpha \beta$ block if layer $\alpha$ and $\beta$ are connected, and \textbf{0} otherwise~\cite{kivela14}. An example of supra-adjacency matrix of a multiplex network is
\begin{equation}
    \label{eq:supra-adjacency-matrix}
    \mathcal{A} = 
        \left(
            \begin{array}{cccc}
                \text{\textbf{A}}^{(1)} & \text{\textbf{I}} & 0 & \cdots \\
                \text{\textbf{I}} & \text{\textbf{A}}^{(2)} & \text{\textbf{I}} & \cdots \\ 
                0 & \text{\textbf{I}} & \text{\textbf{A}}^{(3)} & \cdots \\
                \vdots & \vdots & \vdots & \ddots 
            \end{array}
        \right)
\end{equation}

\noindent
This corresponds to a particular case in which only adjacent layers are connected. The advantage of this second formulation is that it is intuitive and allows the direct application of many single-layer network analysis tools to the larger supra-network.

Following these representations, we can also define the corresponding \textit{Laplacian} matrices. For a single-layer graph, the \textit{Laplacian matrix} is defined as $\text{\textbf{L}} = \text{\textbf{D}} - \text{\textbf{A}}$, where $\text{\textbf{D}}$ is the diagonal matrix of node degrees. Similarly, the \textit{supra-Laplacian matrix} $\mathcal{L}$ can be constructed from the supra-adjacency matrix $\mathcal{A}$~\cite{kivela14, gomez2013}. While the supra-adjacency matrix tells us which connections exist and how they are organized, the supra-Laplacian goes a step further; it captures how processes take place along those connections. 
It extends the analysis from merely identifying the links to understanding how different elements within the network influence each other. This transition from the static structure of connections to the dynamic interplay of influences is pivotal for understanding the overall behaviour of multiplex networks, setting the stage for a deeper exploration of the network's properties through its eigenvalues~\cite{sole2013}.

\subsubsection{Structural Metrics}
\label{sec:metrics}

Once a multilayer network is defined, we need a toolkit of metrics to quantify its structure. While many metrics are generalizations from the ones in single-layer networks~\cite{Aleta2020Nov}, the most insightful ones are those that capture the unique features arising from the multi-layered structure (see Section \ref{sec:multimetrics}).

Starting with node-level metrics, the most basic measure of a node's importance is its \textit{degree}, $k$, defined as the number of connections it has. In multilayer networks, a node $i$ may have a different number of connections in each layer $\alpha$, yielding the per-layer degree $k_i^{\alpha}$. The \textit{multi-degree} of a node is then the sum of its degrees across all layers. The degree is also a simple measure of the \textit{centrality} of a node, in that it expresses its importance in the network. Over the last two decades, the list of centrality measures has considerably expanded, and each one of these metrics is defined according to some rules that limit their range of validity and interpretability. Multilayer networks are well-equipped with an extensive list of centrality measures, many of them being generalizations of the corresponding definitions in single-layer networks. Some of them are purely topological, such as the eigenvector, the Katz, and the betweenness centralities~\cite{DeDomenico2013, DeDomenico2015ranking, sola2013eigenvector, taylor2021}. Others are defined by means of dynamical processes. Notable examples include PageRank~\cite{halu2013multiplex} or a centrality related to the random walk occupation probability~\cite{sole2016random}, among others.

\subsubsection{Spectral properties}
\label{sec:spectral}

To characterise the network as a whole, spectral properties of the system’s matrix representations have been widely used. The eigenvalues of the Laplacian matrix of a network offer a wealth of information, which can be utilized to understand various dynamic properties of the network~\cite{arenas2006, Almendral2007}. The \textit{algebraic connectivity} or \textit{Fiedler value}~\cite{Fiedler1973}, given by the second-smallest eigenvalue ($\lambda_2$) of the supra-Laplacian matrix, is a key metric. A larger $\lambda_2$ indicates a more cohesive and robust network that is harder to break apart. Conversely, a small $\lambda_2$ can indicate potential points of vulnerability or fragmentation within the network. Higher eigenvalues $ \lbrace \lambda_3, \lambda_4, \,\ldots \rbrace $ offer insights into more complex network structures and dynamics\cite{sole2013}. For instance, they can be used to detect community structure within the network~\cite{chauhan2009} (see Section \ref{sec:CD}).

\begin{figure}
\includegraphics[width=0.45\textwidth,clip=true]{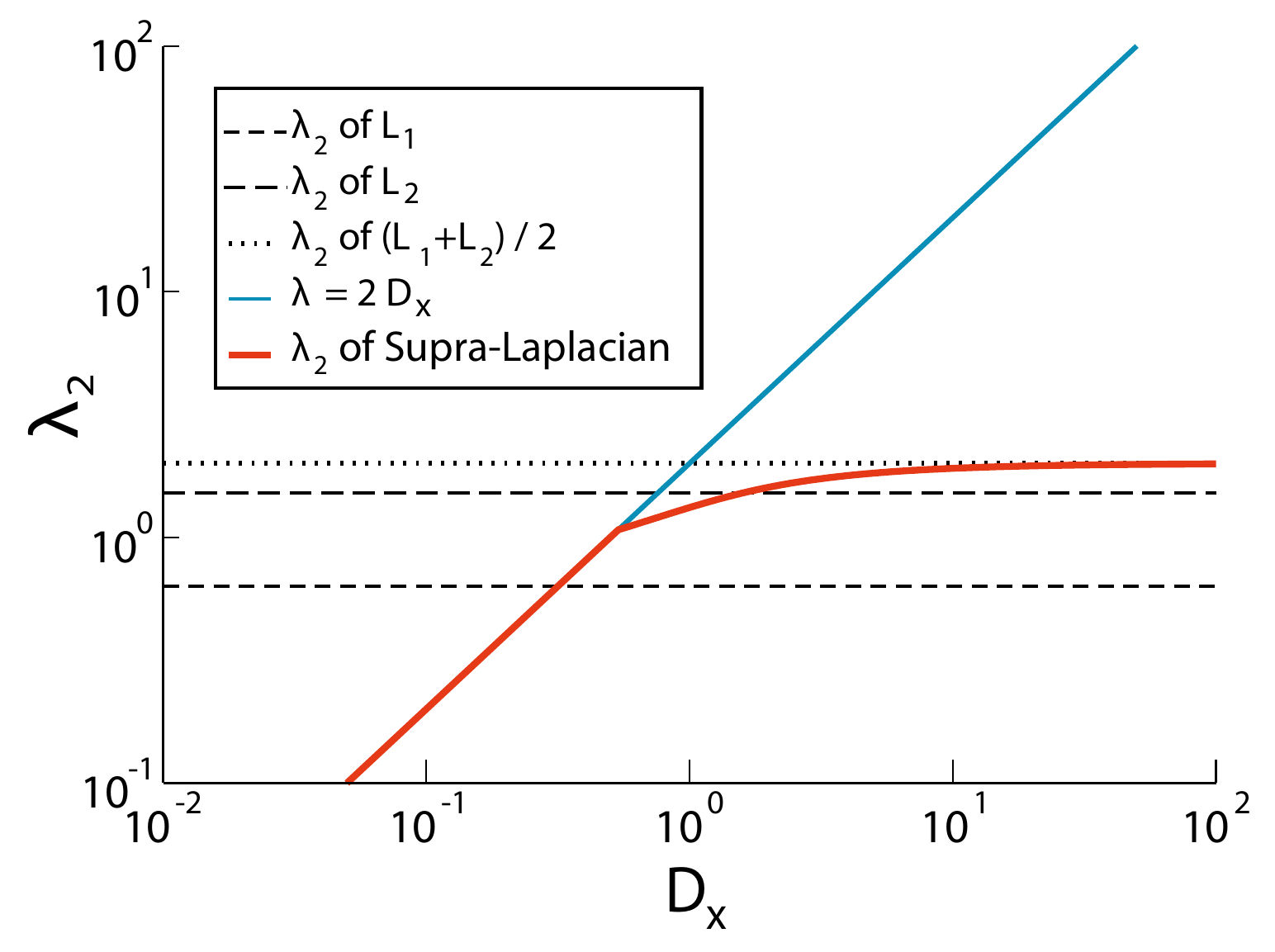}
\caption{Comparison between the second smallest eigenvalues $\lambda_2$ of the different Laplacian matrices as a function of the interlayer coupling $D_x$. Reprinted figure with permission from~\cite{gomez2013}. \copyright 2013 by the American Physical Society.}
\label{fig:laplacians}
\end{figure}

Beyond structural insight, $\lambda_2$ plays a fundamental role in dynamical processes such as diffusion, consensus, and especially synchronisation. In networks where each node represents a Kuramoto phase oscillator, $\lambda_2$ determines the timescale for convergence to the synchronised state~\cite{Almendral2007}. Small values of $\lambda_2$ are associated with bottlenecks or weakly connected components that can delay full synchronisation.
This spectral framework naturally extends to \textit{multiplex networks}, where each node is present in multiple layers.
Figure~\ref{fig:laplacians} shows the theoretical predictions for $\lambda_2$ of a two-layer network in the asymptotic regimes: $\lambda_2 = 2D_x$ for $D_x \ll 1$, and the superposition of both layers for $D_x \gg 1$, $D_x$ being the interlayer coupling. The plot also includes the eigenvalues of the individual layers and the corresponding supra-Laplacian eigenvalue.

Actually, through the behaviour of $\lambda_2$ as a function of the interlayer connectivity, we can directly identify when multiplex networks will behave effectively as if each layer were an independent network, or instead when multiplexity kicks in with its emerging effects on the dynamics. This has been shown in a simplified model of a two-layered network where the interlayer connections are weighted by a homogeneous parameter $p$ in~\cite{radicchi2013abrupt}, but the arguments are general for networks with a larger number of layers. The algebraic connectivity of the supra-Laplacian $\mathcal{L}$ reads $\lambda_2(\mathcal{L}) = 2p$ for $p<p^*$ and is upper-bounded such that $\lambda_2(\mathcal{L}) \leq \lambda_2 ( \mathcal{L_A} + \mathcal{L_B})/ 2$ for $p \geq p^*$, the critical point being $p^* \leq \lambda_2 ( \mathcal{L_A} + \mathcal{L_B})/ 4$. As it can be guessed graphically, but proved mathematically, the first derivative of $\lambda_2(\mathcal{L})$ is discontinuous. This emerges as a result of the crossing of two distinct groups of eigenvalues. Actually, these crossing events appear for the other eigenpairs of the graph Laplacian, with the exception of the smallest and largest eigenvalues. Consequently, this results in discontinuities in the first derivatives of the corresponding eigenvalues~\cite{radicchi2013abrupt}.

The stability of the synchronised state in these systems can be analysed using the Master Stability Function (MSF) formalism~\cite{pecora98}, which allows us to reduce the problem to a spectral one.
The synchronisability of a network can be assessed by the eigenratio
\begin{equation}
    R = \frac{\lambda_N}{\lambda_2},
\end{equation}
where $\lambda_N$ is the largest eigenvalue of the Laplacian matrix. A smaller eigenratio $R$ implies greater synchronisability, since it indicates a wider range of coupling strengths for which the synchronised solution remains stable.

\begin{figure}[ht]
    \centering
    \includegraphics[width=0.45\textwidth]{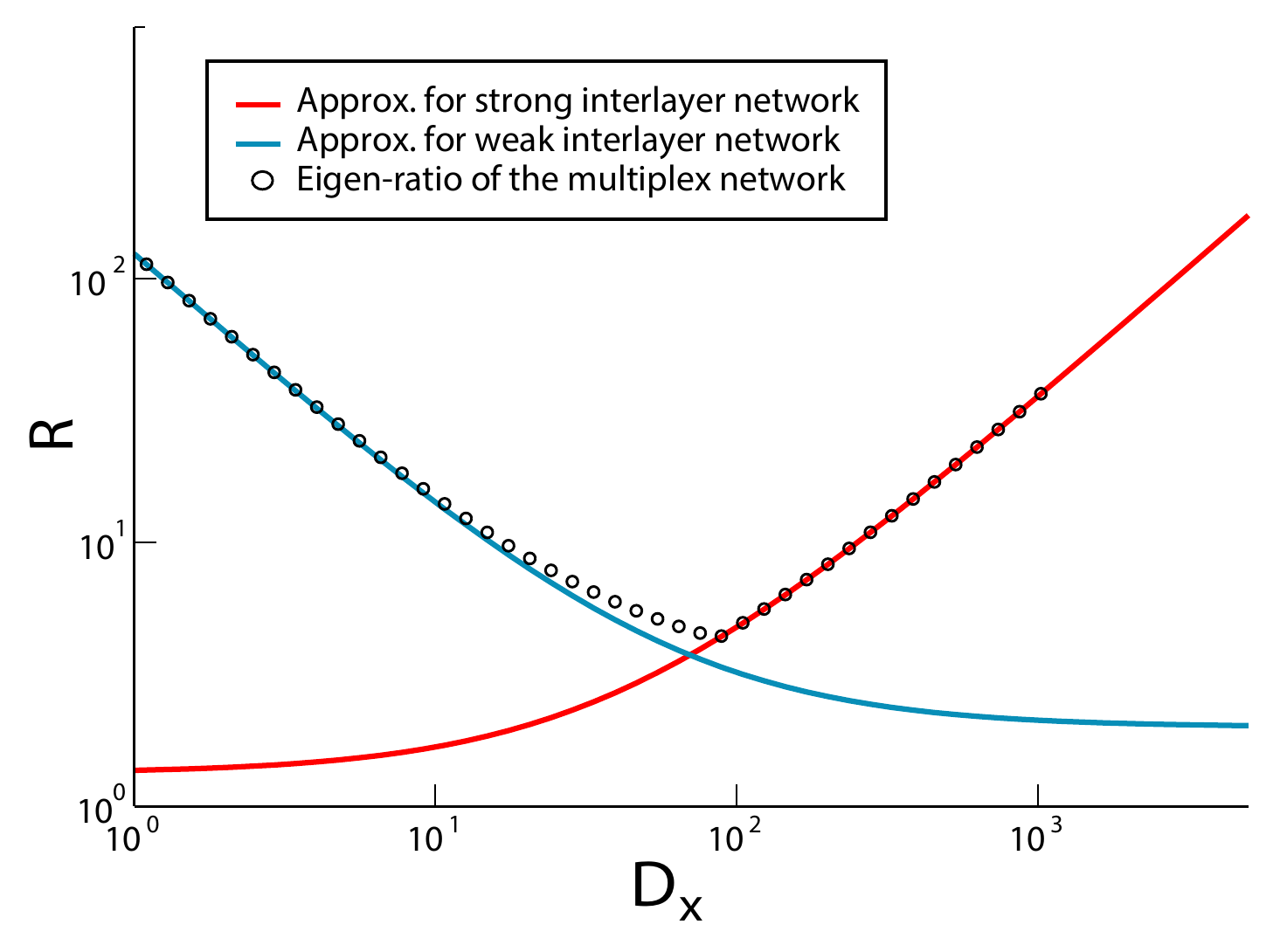}
    \caption{Eigenratio $R = \lambda_N / \lambda_2$ as a function of the interlayer coupling strength $D_x$. Circles represent exact numerical results, while the solid lines show the analytical approximations valid in the weak- and strong-coupling limits. Adapted figure with permission from~\cite{sole2013}. \copyright 2013 by the American Physical Society.}
    \label{fig:eigenratio_approx}
\end{figure}

In~\cite{sole2013}, analytical approximations for the eigenratio $R$ were provided in two limiting regimes: weak interlayer coupling ($D_x \ll 1$) and strong interlayer coupling ($D_x \gg 1$). These approximations offer explicit insight into how the topology and interlayer coupling influence the dynamics. The predictions were shown to agree well with numerical results in a wide range of coupling values, as illustrated in Figure~\ref{fig:eigenratio_approx}. The point at which the two analytical regimes intersect provides an estimate of the optimal coupling strength for synchronisation performance.

These results support the use of spectral indicators, particularly $\lambda_2$ and the eigenratio $R$, as powerful tools for connecting structure and dynamics in multilayer networks.

\subsubsection{Multilayer metrics}
\label{sec:multimetrics}

Some metrics and concepts do not have counterparts in standard single-layer networks. The characterisation of multilayer networks through these new metrics has unveiled an array of phenomena hitherto unobserved on single-layered systems, thus manifesting the singularity and usefulness of the multilayer description. For instance, the \textit{participation coefficient} measures how evenly a node's links are spread across the different layers. A node with high multi-degree but low participation coefficient is a \textit{hub} - a highly connected node - in only one or a few layers, while a node with high participation coefficient is genuinely active across the entire system. This distinction is crucial for understanding a node's role in processes that span multiple layers~\cite{battiston14}.

Another network-level metric with no single-layer equivalent is the \textit{mutually largest connected component} (MLCC). This is the largest group of nodes where every node can reach every other node through paths existing within each and every layer~\cite{buldyrev_2010}. The size of the MLCC is a powerful indicator of a system's integrated resilience~\cite{artime2024robustness}. In many multilayer systems, reducing network connectivity (e.g., by removing nodes or links) can cause the MLCC to collapse abruptly in a discontinuous phase transition, contrasting sharply with the smoother, continuous transitions typically seen in single-layer networks~\cite{gao2012networks, radicchi2013abrupt}.

A fundamental question in multilayer network science is: ``What new information does the layered structure provide compared to a simple, aggregated network?'' To answer this, we need metrics that quantify the relationships between layers. One approach is to measure layer similarity or redundancy. This can be done by calculating the interlayer mutual information or other correlation measures between the adjacency matrices of different layers. If layers are highly similar, the system is redundant, and an aggregated view might suffice. If they are diverse, the multilayer representation is essential, as different layers capture distinct relational patterns~\cite{DeDomenico2015ranking,Schieber2017,ghavasieh2020enhancing}.

Furthermore, multilayer networks introduce new types of structural correlations. Beyond intralayer patterns like degree assortativity (whether high-degree nodes connect to other high-degree nodes~\cite{pastor01b,newman02}), we can measure interlayer correlations. For example, are nodes with a high degree in one layer also likely to have a high degree in another? The presence or absence of such correlations significantly impacts the behaviour of dynamical processes, from spreading to synchronisation, and has been a major focus of research~\cite{pamfil2019relating, nicosia2015measuring, wu2020correlated, nicosia2013growing, bianconi2013statistical, battiston14, menichetti2014weighted, lee2012correlated}. 

\subsection{Structure and Dynamics\label{sec:structure_dynamics}}

\subsubsection{Community detection\label{sec:CD}}

How shall we define communities in multilayer networks? The answer is context-dependent. In a network of customers layered by product categories, a community might be a group of people with consistently similar purchasing habits across multiple layers (a cross-market segment). In a biomedical context, coupling a gene co-expression network with a protein-protein interaction network, a community could reveal groups of genes that are co-expressed and whose encoded proteins physically interact. The search for such meaningful groups in multilayer networks has become a vibrant research area, with several reviews already charting the landscape~\cite{kim15,magnani21,huang21}. Here, we summarize the main strategies and highlight key challenges and future directions.

Community detection is mostly performed on multiplex networks, where the same node exists across all layers. This framework is also directly applicable to temporal (multi-slice) networks, treating each time snapshot as a layer (see Section \ref{sec:temporal}). Methodologically, approaches to find multilayer communities can be grouped into three main strategies: flattening, layer-by-layer, and multilayer.

The first strategy, flattening, collapses the multilayer network into a single aggregated graph containing all nodes and links from all layers~\cite{berlingerio11,kim16}. Any traditional algorithm working on single-layer networks can then be applied. While computationally efficient, this approach discards all information about the distinct layers and their relationships.

Layer-by-layer algorithms first detect communities on each layer independently and then aggregate the results. These are further divided into three classes: frequent subgraph mining~\cite{zeng06,berlingerio13}, feature integration~\cite{dong13,nie18}, and partition integration~\cite{cantini15,tagarelli17}. Frequent subgraph mining is based on the idea that the nodes of a multilayer community are assigned to the same group in different layers. Thus, it relies on identifying groups of nodes that are consistently clustered together across many layers. Feature integration relies on the principle that communities are made of nodes that are similar to each other, something typically measured by looking at the eigenvectors of the layers. Communities are found by grouping nodes according to their features via, e.g., k-means clustering~\cite{forgy65,macqueen67}. Partition integration directly combines the divisions found on the individual layers. For instance, one could derive a consensus partition, which is the division that is closest to all the input ones, on average~\cite{strehl02}.

\begin{figure*}
\begin{center}
\includegraphics[width=\textwidth]{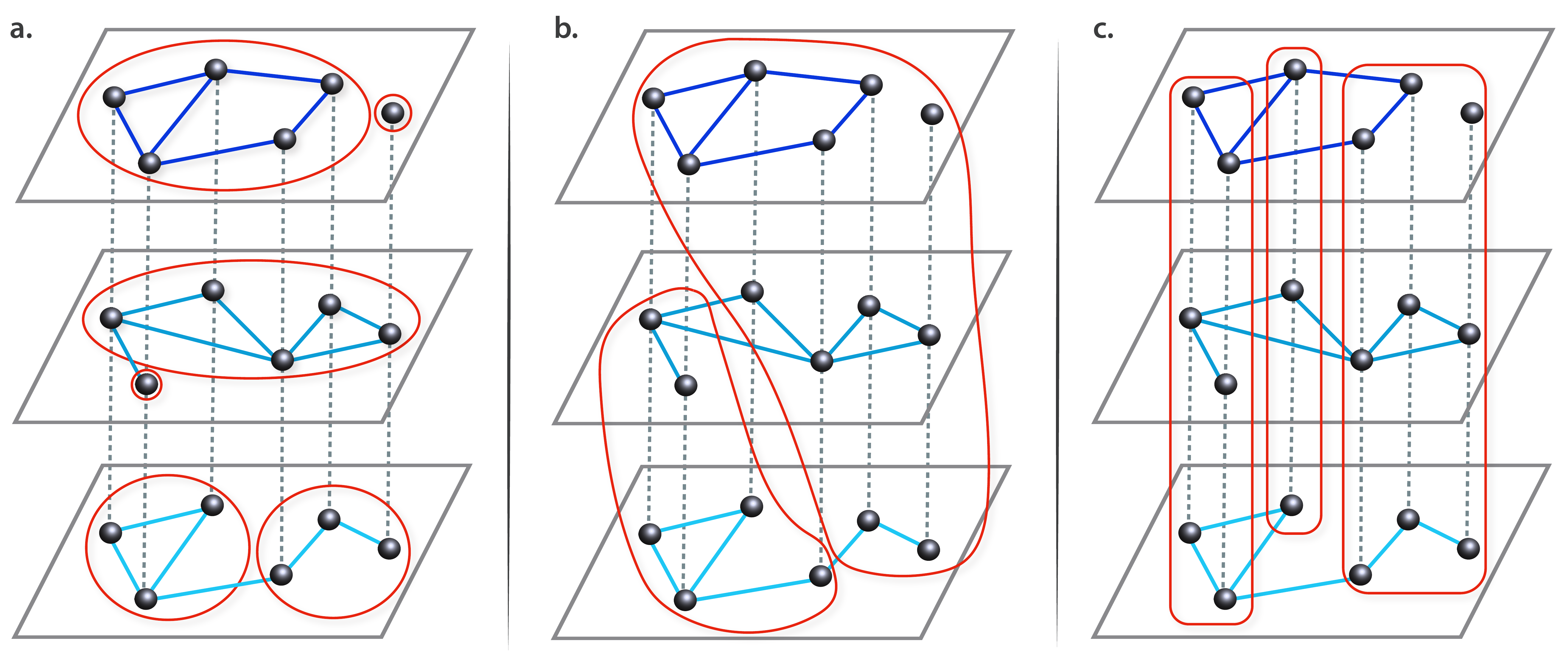}
\caption{Multilayer communities. The continuous (red) contours indicate communities. (a) Intralayer communities group state nodes on the same layer. (b) Members of cross-layer communities can be on both the same and different layers.
(c) Pillar communities group all the counterparts of the nodes in other layers in the same community.}
\label{fig:multilayer_comml}
\end{center}
\end{figure*}

Lastly, we have genuinely multilayer methods. These algorithms operate on the complete multilayer structure without prior aggregation. Furthermore, they do not cluster single nodes but state nodes (or node-layer tuples, i.e., a specific node on a specific layer). This allows for richer community definitions (Fig.~\ref{fig:multilayer_comml}):
\begin{itemize}
    \item Intralayer communities group state nodes on the same layer.
    \item Cross-layer communities can group state nodes from different layers.
    \item Pillar communities are a constrained type where if a node belongs to a community, all of its counterparts in other layers also belong to the same community.
\end{itemize}

Many of these methods are a direct extension of single-layer algorithms to the multilayer case. Modularity maximization was generalized by defining a multilayer null model based on random walks, with intralinks and interlinks playing different roles~\cite{mucha10}. Local exploration methods, like seed set expansion~\cite{interdonato17}, label propagation~\cite{boutemine17} or the clique percolation method~\cite{afsarmanesh18}, as well as random walk-based algorithms such as Infomap~\cite{dedomenico2015identifying} and Walktrap~\cite{kuncheva15}, have also been adapted. Furthermore, statistical inference methods, primarily based on the stochastic block model (SBM), offer a principled framework to detect communities, infer relationships between layers, and even test whether an observed single-layer network is better explained as an aggregation of hidden layers~\cite{peixoto15,debacco17}.
				
Despite this progress, significant challenges remain. First, a persistent issue is the rigorous assessment of statistical significance. Without it, one cannot distinguish meaningful communities from patterns arising by chance. Second, the increasing availability of node and edge metadata is often used only for validating results, a procedure known to be unreliable~\cite{hric14,peel17}. Instead, this information could be directly integrated into the detection process to find more accurate and relevant communities, a task for which SBM-based approaches are particularly well-suited~\cite{hric16,newman16}. Finally, the vast majority of methods are designed for multiplex networks. There is a clear need for algorithms that can handle more general multilayer networks (often called heterogeneous networks in computer science) with different node sets and relationships in each layer, a promising direction for future research~\cite{sun13}.

\subsubsection{Dynamics and Processes}
\label{sec:dynamics}

The multiplex network framework allows for a wide variety of dynamical behaviours that go beyond simple extensions of single-layer dynamics. Unlike isolated networks, where each node interacts through a single set of connections, multiplex networks allow each node to participate in multiple interaction layers, each possibly governed by different dynamics, rules, or time scales. This structure enables novel emergent behaviours that are fundamentally rooted in the interplay between dynamics on and across layers.

In some fields, particularly computer science, it is common to distinguish between \textit{actors} (the underlying real-world entities) and \textit{nodes} (their representations within specific layers)~\cite{czuba2024networkdiffusion, Zhong2022-cp, Czuba2025-oz}. While this specific terminology may be less explicit in other domains, the underlying principle of defining processes at the level of the entity, rather than solely at the layer-specific node, is widely relevant. This framework allows dynamics to be driven by the actor, integrating states from all its corresponding nodes simultaneously, or by specific layer-level copies. Consequently, it enables complex activation rules: an actor might change state only if a threshold of its nodes are active (reinforcement), or conversely, activity in a single node might drive the entire actor (contagion).

In what follows, we review several representative classes of dynamical processes that have been studied in the context of multiplex networks, emphasizing those in which the multilayer structure is essential to the observed phenomena. The discipline-specific applications of these processes will be presented later in Section~\ref{sec:appl}.

A paradigmatic case in which multiplexity alters the behaviour of a well-understood process is that of reaction-diffusion systems. In multiplex networks, reaction and diffusion terms can be assigned to different layers or coupled across them. For instance, in~\cite{Kouvaris2015}, activator and inhibitor species are placed on separate layers, where each species diffuses only within its own layer but reacts across layers. Remarkably, the presence of interlayer coupling allows for Turing-like pattern formation even when both species have identical mobility rates—something impossible in single-layer networks. This form of topology-driven instability, further explored in~\cite{Asllani2014}, highlights how multiplexity can give rise to heterogeneous and robust spatial patterns under conditions where the dynamics of isolated layers remain homogeneous.

In multiplex systems, one well-studied example of interacting dynamics is the coupling between epidemic spreading and awareness diffusion~\cite{Granell2013Sep}. In this model, the disease propagates through physical contact on one layer, while information about the disease (awareness) spreads via virtual communication on a separate layer. The two processes mutually inhibit each other, giving rise to a point at which the presence of awareness fundamentally alters both the epidemic threshold and the eventual prevalence. 

Multiplex networks also provide a natural setting for studying feedback between processes on different layers. In~\cite{Nicosia2017}, a model is introduced in which neural synchronisation on one layer modulates resource allocation on another, and vice versa. This mutual feedback leads to the spontaneous emergence of explosive synchronisation and highly heterogeneous resource distributions.

Multiplex networks often feature layers with different topologies or dynamics, leading to hybrid systems. In~\cite{Hizanidis2016}, a modular neural network is modelled with electrical synapses within communities and chemical synapses across them, each type represented as a different layer. The interplay between these types of couplings gives rise to chimera-like states, where coherent and incoherent activity coexist across modules. Such hybrid multiplex models are relevant to understanding how local synchrony coexists with global disorder in brain dynamics.

A similar approach is taken in~\cite{Sadilek2015}, where a two-layer Kuramoto model is derived from physiologically motivated equations for cortical dynamics. The model captures transitions between resting and background brain activity, successfully reproducing empirical features that were inaccessible to simpler models. Again, multiplexity allows for distinct coupling schemes across layers (e.g., electrical vs chemical, or fast vs slow processes) that shape the global phase diagram.

Another work examined how two layers not directly connected can synchronise through the mediation of a central relay layer~\cite{Leyva2018}. Results showed that synchronisation thresholds are reduced in the multiplex setting and that low-degree nodes play a crucial role, while hubs can be “demultiplexed” without disrupting global coherence. More recently, a three-layer multiplex of non-linear oscillators, in which a central drive layer unidirectionally couples to two response layers, has been investigated~\cite{Vadakkan2025}. Such directional coupling induces synchronised oscillations in the response layers, even with amplification relative to the drive, and varying interlayer strength or introducing time-scale mismatches enables control over synchronisation patterns, amplitudes, and dynamical regimes such as quasi-periodicity and generalized synchronisation.

These examples collectively illustrate that multiplex networks enable a new class of dynamical processes: systems where different types of dynamics not only coexist but interact, often through cross-layer mechanisms. This interplay can give rise to topology-induced instabilities, extended metastable states, abrupt phase transitions, novel critical points, and even the coexistence of order and disorder. 

Such behaviours are not mere quantitative variations of single-layer dynamics but emerge qualitatively due to the multiplex architecture. Importantly, these phenomena are often robust to parameter variations and are observed across different domains, from neuroscience and epidemiology to social systems. In Section \ref{sec:appl}, we will revisit some of these dynamical mechanisms in the context of concrete applications, where domain-specific constraints and empirical data can further shape the interaction between structure and dynamics.

\subsection{Advanced Frameworks\label{sec:advanced}}

\subsubsection{Temporal Networks\label{sec:temporal}}

Complex systems, from social groups to interacting neurons, are fundamentally dynamic. While classic network science often represents these systems with static graphs, this is an aggregation that loses all temporal information, such as the sequence, duration, or concurrency of interactions. The recent availability of time-resolved data and computational power has propelled the study of temporal networks, revealing phenomena invisible in a static view~\cite{holme2012temporal,holme2015modern}. A temporal network can be represented as a sequence of static snapshots, which naturally fits the multilayer network framework where each layer is a snapshot in time. Besides, it is also possible to have networks that are temporal and multilayer at the same time, in which case the temporal correlations between layers can significantly affect the dynamics of processes unfolding on multilayer networks \cite{starnini2017effects}.

\begin{figure*}
\begin{center}
\includegraphics[width=\textwidth]{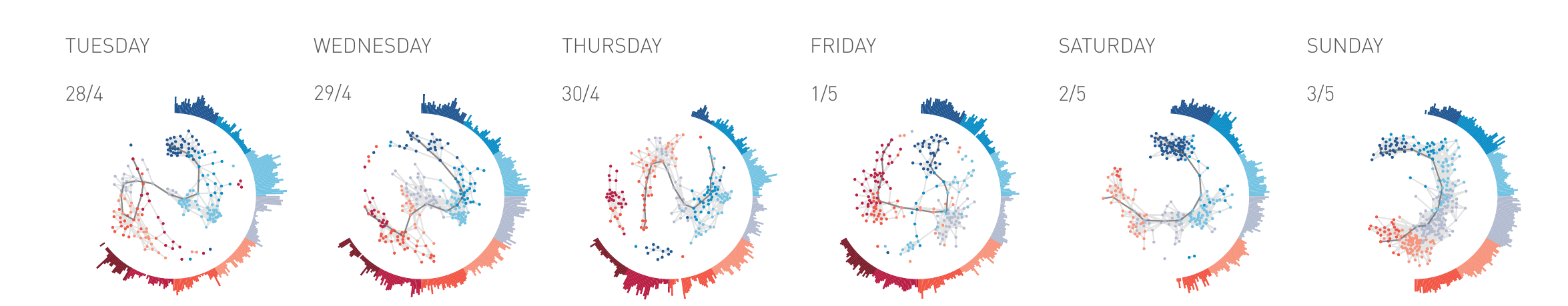}
\caption{Temporal networks of face-to-face interactions of people on different days of museum visits. In each day, time evolution is represented as a colour code. The data has been recorded by detecting physical proximity using wearable sensors developed by the SocioPatterns collaboration \cite{isella2011what,vandenbroeck2012making}. Reprinted figure with permission from the SocioPatterns collaboration.}
\label{fig:temp}
\end{center}
\end{figure*}

While early theoretical models added temporality by dynamically changing links, particularly in an adaptive fashion \cite{gross2008adaptive}, the strong expansion of the field has been mostly data-driven~\cite{holme2012temporal,holme2015modern}. The availability of high-resolution temporal data, such as the face-to-face contact network in a museum exhibition depicted in Figure~\ref{fig:temp} and collected by the SocioPatterns collaboration using wearable sensors~\cite{cattuto2010dynamics}, provided a strong empirical foundation for the field. This data made it clear that static representations not only hide temporal details but can be misleading, as the properties of a networked system can vary dramatically from one moment to the next~\cite{masuda2016guide,saramaki2015seconds}.

This data-centric view is what led researchers to analyse temporal networks as a series of snapshots. This approach revealed that while aggregated properties might converge to the static ones~\cite{krings2012effects}, local measures like a node's centrality could fluctuate significantly, with nodes shifting from central to peripheral roles over time~\cite{braha2006centrality,braha2009time}.  However, a true paradigm shift occurred when the focus moved from fluctuating static properties to inherently temporal concepts. Researchers recognized that information flows along temporal paths, which are a set of time-respecting sequences of interactions~\cite{kempe2000connectivity,holme2005network}. A path that exists in an aggregated static network may be impossible in a temporal one if the interactions do not occur in the correct chronological order~\cite{pan2011path}. This insight spurred the redefinition of classical centrality measures for the temporal domain~\cite{stehle2010dynamical} and led to the discovery of new phenomena, such as the burstiness of human interactions, where events are clustered in time rather than occurring randomly~\cite{barabasi2005origin,karsai2018bursty}.

But not all temporal interactions are equally informative. Patterns may appear briefly, recur, or vanish entirely. Motifs that are evident in aggregated static views of the network may never exist in real time due to improper chronological alignment~\cite{holme2005network}. To deal with this complexity, researchers have developed various techniques for filtering and pattern detection. At the local scale, temporal backbones highlight node pairs whose interaction frequencies significantly exceed what would be expected under a temporal null model~\cite{kobayashi2019structured}. Temporal motifs identify frequently recurring, time-ordered subgraphs~\cite{kovanen2013temporal}, while temporal cores generalize the static core decomposition to uncover dense substructures active over defined time windows~\cite{galimberti2018mining}. Similarly, the temporal rich-club coefficient measures whether highly connected nodes in the aggregate network tend to form simultaneous, persistent connections~\cite{pedreschi2022temporal}.

In parallel, significant effort has gone to identifying the relevant timescales that govern the evolution of temporal networks. Early techniques relied on comparing successive time windows~\cite{darst2016detection}, but this overlooked long-range correlations. More sophisticated methods compute pairwise distances or similarities between network snapshots~\cite{masuda2019detecting}, enabling the clustering of recurring states and the identification of their transitions. Other approaches apply auto-correlation metrics, or Fourier transforms to extract temporal correlations and recurrent interaction patterns at different time scales~\cite{lacasa2022correlations,andres2024detecting}.

Representing temporal networks in a simple form for analysis has also been a priority. Two lossless, static-like representations have emerged as especially useful. The supra-adjacency representation encodes node-time pairs as nodes in a multilayer structure, linking them according to the observed interactions~\cite{valdano2015analytical,sato2019dyane}. The event graph, in contrast, treats interactions themselves as nodes, linking temporally adjacent events involving shared participants~\cite{kivela2018mapping}. This latter framework is especially well-suited to model processes constrained by timing, such as epidemic spreading or transit dynamics~\cite{badie2022directed}. Another line of work uses graph embedding techniques to map high-dimensional temporal networks into low-dimensional vector spaces~\cite{goyal18,cai18}. These embeddings preserve structural or temporal similarities and support tasks such as link prediction or influence sets \cite{thongprayoon2023embedding, torricelli2020weg2vec,dall2024embedding}. Embeddings also facilitate data compression and learning-based applications, as discussed further in Section~\ref{sec:ML}.

Models of temporal networks have been developed to generate realistic synthetic networks or to establish null benchmarks. Mechanistic models aim to reproduce observed features or test hypotheses about dynamics~\cite{holme2015mechanistic}, while randomized models offer a controlled way to destroy correlations and assess their significance~\cite{gauvin2022randomized}. These models range from memoryless activity-driven frameworks~\cite{perra_activity_2012,starnini_topological_2013} to non-Markovian extensions capturing burstiness and memory effects~\cite{moinet_burstiness_2015}. Additional approaches include stochastic block models, priority queue models, models based on Hawkes or other dynamical processes, and memory network models. 

The temporal networks framework also adds another layer of complexity to dynamical processes unfolding on them, since one needs to take into account the coupling between the characteristic timescale of the dynamic process and the temporal network's one. In epidemiology, the coupling between interactions and disease progression profoundly affects spreading outcomes with important consequences for designing effective public health interventions~\cite{masuda2017introduction,machens2013infectious, koher2019contact, toroczkai2007proximity, valdano2015analytical}. In social systems, temporal networks are used to model the spread of information, behaviours, and innovations, often through threshold mechanisms distinct from disease transmission~\cite{unicomb2021dynamics,andres2024competition}. Temporal opinion dynamics have been explored to understand consensus and polarization~\cite{li2019impact,chu2023non}, while synchronisation phenomena in biology and engineering are modelled as coupled oscillators interacting over time-varying links~\cite{ghosh2022synchronized}. Applications in transportation include traffic optimisation and infrastructure design~\cite{salama2022temporal}.

Looking forward, a central challenge is identifying the minimal yet sufficient temporal structures necessary for specific applications. For example, while the temporal paths in a physical proximity network may code the ways an epidemic would spread between the people, several interactions may not fall on these paths and thus appear completely irrelevant from the aspect of a spreading process. Thus, since many interactions may not influence a process, reducing networks to effective backbones could lead to smaller and faster computations, as well as improving data sharing in a privacy-preserving way. 

\subsubsection{Higher-order interactions\label{sec:higher}}


Group interactions are widespread in real-world systems, ranging from social dynamics~\cite{Iacopini2019,Arruda2020,Battiston2021,Landry2020,St-Onge2022,Arruda2023,Kiss2023}, such as group chats, to chemical reactions involving multiple reactants~\cite{Jost2019,Battiston2021,Traversa2023}. In these contexts, the traditional network science framework, which considers only pairwise interactions, is often insufficient to capture the complex dependencies between multiple interacting entities. Higher-order models provide a richer framework for accurately representing these systems and avoiding misleading predictions~\cite{Battiston2020,Battiston2021,Arruda2023,FerrazdeArruda2024Aug}. For example, contagion processes that model the population as a graph will present a second-order phase transition~\cite{Pastor-Satorras2015,Arruda2018}. However, contagion models in higher-order networks often exhibit discontinuous transitions~\cite{Iacopini2019, Arruda2020, Landry2020, St-Onge2022, Arruda2023, FerrazdeArruda2024Aug}. Similarly, in pairwise contagion models, one expects a single stable solution, while multistability is possible in contagion models in higher-order networks~\cite{Arruda2023,Kiss2023,FerrazdeArruda2024Aug}. Interestingly, these new phenomena are not restricted to contagion processes; they have also been observed in synchronisation processes~\cite{Skardal2023} and evolutionary games~\cite{civilini24}. These examples demonstrate how neglecting the higher-order nature of a process can alter the outcome of a model, potentially leading to incorrect predictions.

Mathematically, higher-order interactions are usually modelled using hypergraphs or simplicial complexes~\cite{Battiston2020,Boccaletti2023,Torres2021}. Hypergraphs allow edges (hyperedges) to connect any number of nodes, whereas simplicial complexes require that all subsets of a hyperedge be present. The best choice of representation often depends on the dynamics being studied. A discussion about this topic can be found in~\cite{Torres2021}.

\begin{figure}[ht]
    \centering
    \includegraphics[width=0.45\textwidth]{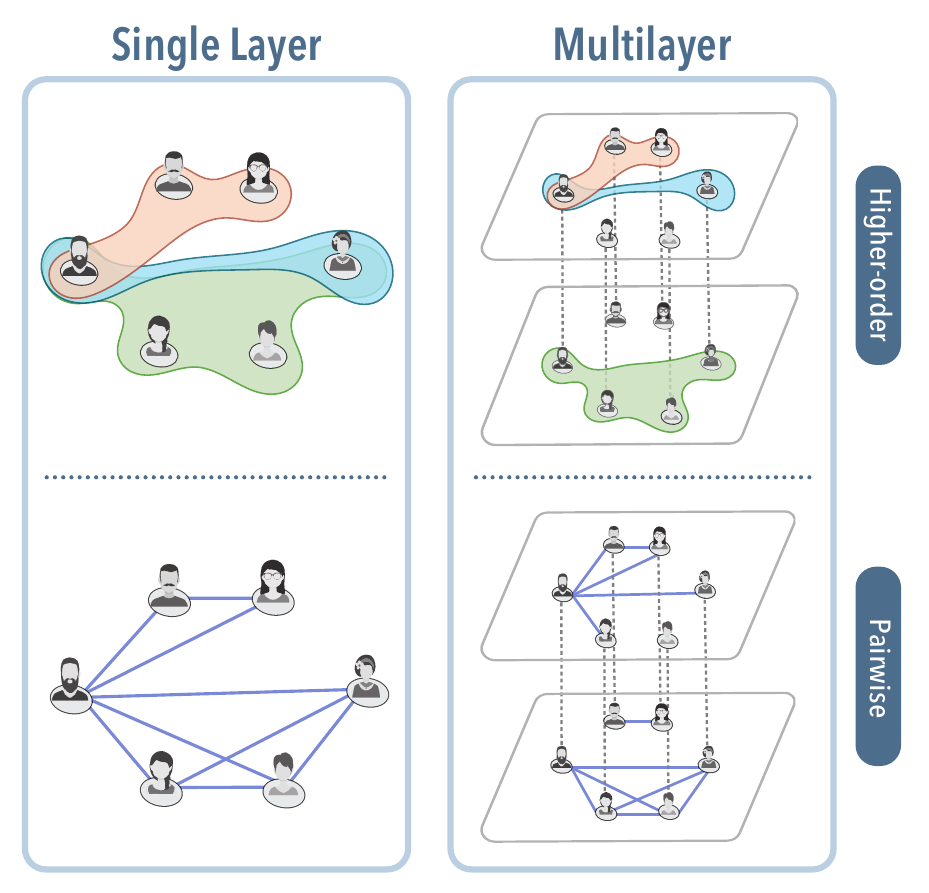}
    \caption{Schematic representation of different networked systems. The bottom row depicts systems with pairwise interactions, while the top row contains group interactions including more than two nodes. As with classical pairwise models, it is possible to associate a type to each interaction, extending the framework of multilayer networks to multilayer hypergraphs. Adapted figure with permission from~\cite{FerrazdeArruda2024Aug}. \copyright 2024 by Springer Nature.}
    \label{fig:hypergraph}
\end{figure}

Compared to pairwise networks, the structural and spectral analysis of higher-order systems is still in development. While concepts such as adjacency and Laplacians are well-defined in networks, their hypergraph counterparts are more complex and often combine structural and dynamical parameters~\cite{Schaub2020,Mulas2020,Arruda2021}. This complicates the analytical study of processes such as diffusion and synchronisation, and different definitions may yield different outcomes depending on the context. Similarly, most inference methods in the literature deal with inferring pairwise relationships. Detecting higher-order structures requires new methods~\cite{Peixoto2020,Contisciani2022}.

Recent work has begun to explore multilayer extensions of higher-order systems (Fig.~\ref{fig:hypergraph}), such as triadic percolation processes and interdependent dynamics across layers~\cite{Sun2023,Chang2023}. 
In the context of spreading processes, a two-layer approach considering the concurrent spread of information and disease was studied in~\cite {fan_epidemics_2022, hong_coupled_2023}. Furthermore, the diffusion of resources coupled with epidemic spreading was examined in ~\cite{sun_diffusion_2022}. These studies demonstrate how combining multilayer and higher-order structures can lead to novel, emergent behaviours, such as oscillatory or chaotic dynamics. Another promising research direction is studying temporal hypergraphs, which extend these representations to time-varying settings. This framework captures interactions involving groups of nodes that occur over finite durations. Examples include simultaneous face-to-face contacts~\cite{di_gaetano_percolation_2024,gallo24b} and group-level dynamics in social settings~\cite{Iacopini2024Aug}. Temporal hypergraphs provide a natural language to describe and model these evolving group structures, opening new avenues for analysing their evolution and how they influence dynamical processes.

\subsubsection{Machine learning}
\label{sec:ML}

Graph representation learning, or graph embedding, aims to map nodes into a geometric space such that structural properties are preserved~\cite{goyal18,cai18}. This task is essential for applying machine learning techniques to relational data and has been widely used for link prediction, node classification, routing, and visualisation. In the past few years, traditional embedding methods have been extended to deal with heterogeneous graphs and knowledge graphs, where nodes or edges carry additional categorical, numerical or textual information~\cite{Shi2022-pi}. Multiplex networks can be seen as a specific type of heterogeneous graph and have prompted the development of dedicated embedding techniques.

A good multiplex embedding method should effectively capture both intralayer and interlayer structures. Multiplex network embeddings can be divided into two major classes: \textit{joint embeddings}, which project all replicas of a node across layers to a single point in a low-dimensional Euclidean space, and \textit{individual embeddings}, where each layer has its own representation of the node. Joint embeddings can be obtained by merging individual embeddings. The general approach followed by most multiplex embedding techniques consists of three phases: sampling, training, and optimisation. 

\begin{figure*}
\begin{center}
\includegraphics[width=0.8\textwidth]{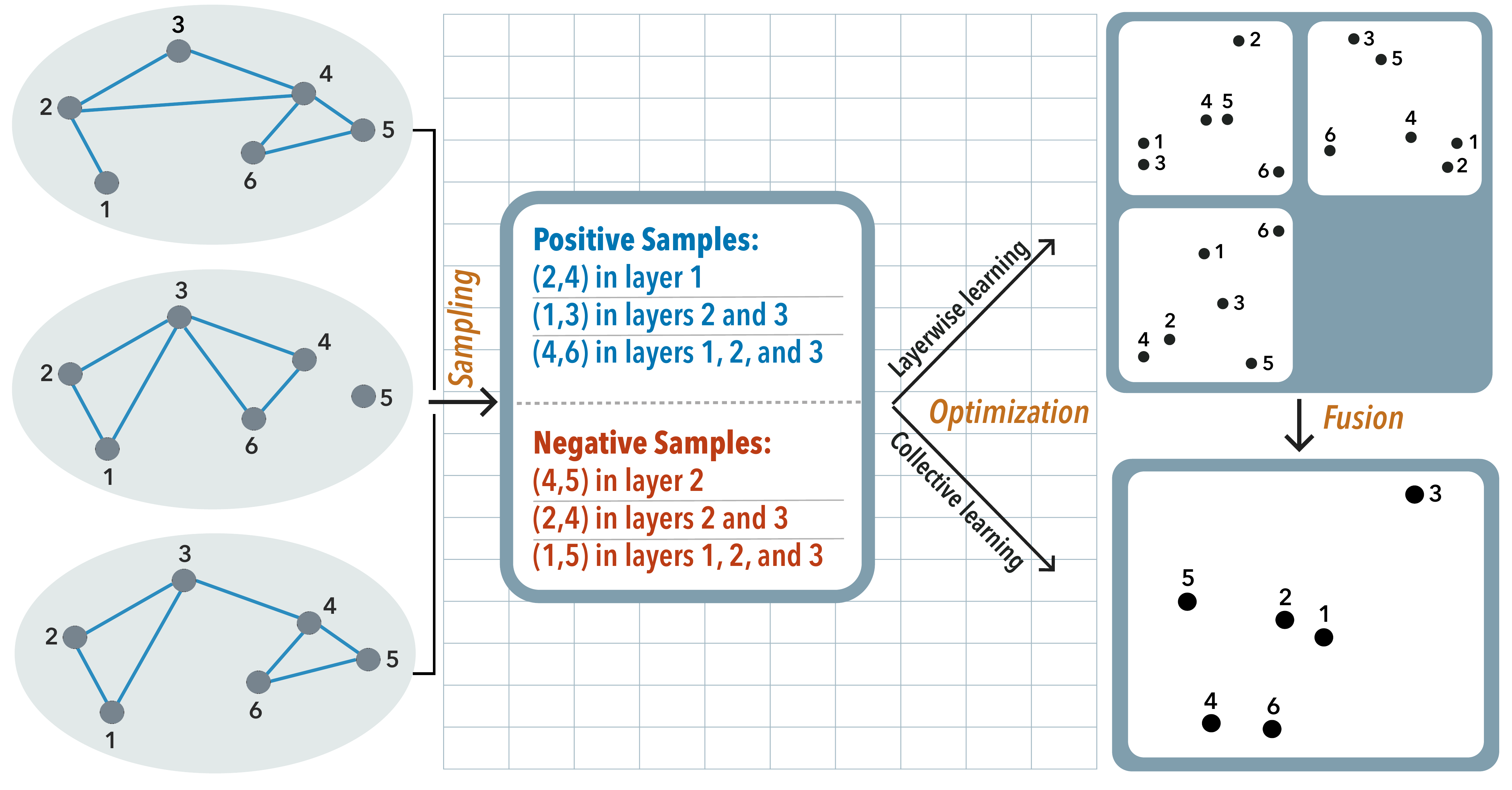}
\caption{General approach to multiplex embedding. Given a multiplex network, the first step involves identifying node pairs that are supposed to have similar (positive samples) and dissimilar (negative samples) representations. Next, the learning framework is used to train the node representations based on the sampled data. The node representations can be learnt for the entire multiplex (bottom right) or for individual layers (top right). Individual embeddings may be fused together to obtain joint embeddings.}
\label{fig:multi_embed}
\end{center}
\end{figure*}

\textit{Sampling}. While embedding, we want pairs of nodes that are close to each other in the network to also have similar embeddings. These are called ``positive node pairs'' and can be found using several strategies. Techniques such as DeepWalk~\cite{perozzi14} and node2vec~\cite{grover16} simulate random walks to define contexts, akin to word embeddings in natural language processing~\cite{mikolov13}. These approaches have been extended to multiplex networks by allowing walks to transition between layers with some probability~\cite{liu17}. Other methods define proximity using K-nearest neighbors or neighborhood-based criteria, as in GraphSAGE~\cite{hamilton17}. To reduce computational cost, negative sampling is used to approximate objective functions, drawing random ``negative pairs'' assumed to be dissimilar~\cite{cai18}.

\textit{Training}. In this phase, multiplex embedding methods learn representations using shallow neural networks, although more advanced architectures have been proposed. In shallow neural networks, the training process is specified by the structural organisation of the learning algorithm, which can be layerwise or collective learning (Fig.~\ref{fig:multi_embed}, top and bottom rows respectively). Layerwise learning separately trains node embeddings for each layer and combines them using weights, often learned via attention mechanisms~\cite{qu17,xie20}. Collective learning directly learns unified representations by leveraging multiplex sampling and skipgram-style optimisations~\cite{zhang18,liu17}. While shallow networks are common, they face important limitations, such as the inability to generalize to unseen nodes or share parameters between nodes (since the encoder optimises a unique embedding vector for each node). To alleviate these limitations, more sophisticated learning architectures, like graph neural networks, have been proposed.

\textit{Optimisation}. Embeddings are adjusted to minimize loss functions that reward proximity between positive pairs and penalize proximity between negative ones. Some methods also include node attributes to improve representation quality~\cite{feng19,cen19}. These optimised embeddings can then be used in downstream tasks such as:
\begin{itemize}
    \item Link prediction: predicting missing or future links within a layer (layer-wise), across layers (interlayer) or in an aggregated view of the multilayer~\cite{liu17,xie20,zhang18,feng19,gallo24}.
    \item Node classification: predicting labels for unlabelled nodes based on known node attributes or partial label data~\cite{zhang18}.
    \item Community detection: discovering modular structures across layers by learning weights that reflect each layer's contribution~\cite{jing21,cai21}.
\end{itemize}

In addition to Euclidean embeddings, hyperbolic embeddings provide a geometry better suited to hierarchical and navigable networks~\cite{garcia19}. These approaches can potentially improve performance in tasks like greedy routing, where shortest paths are found using only local information. Recent work using layerwise hyperbolic embeddings suggests that interlayer geometric correlations improve routing performance~\cite{kleineberg16}. Future work could explore hyperbolic multiplex embeddings more systematically.

Lastly, synthetic multiplex models~\cite{bazzi20} can be used to systematically vary layer-specific properties and interlayer correlations, providing testbeds for evaluating embedding performance under controlled conditions. These models support the study of how properties like modularity and community alignment across layers affect embedding quality and task performance.

\section{Applications}
\label{sec:appl}

Network science has long served as a unifying language across disciplines, offering a flexible framework for modelling complex systems. With the rise of multilayer network theory, this versatility has only expanded. Multilayer networks allow researchers to capture the rich, interconnected structure of systems where multiple types of interactions, temporal dynamics, or functional interdependencies coexist. As a result, their applications span a diverse array of domains—from infrastructure and biology to economics and the social sciences—demonstrating both the theoretical depth and real-world impact of the multilayer perspective.

\subsection{Interdependent systems}
\label{sec:failures}

The importance of multilayer interdependent networks was first recognized by engineers and later formalized through the interdependent percolation model introduced in~\cite{buldyrev_2010}. In addition to standard intralayer links, these systems feature dependency links that connect nodes across layers, such that the functionality of a node in one network depends on the functionality of a node in another. This structure can lead to cascading failures as a breakdown in one layer propagates to others, potentially causing large-scale system collapse. Unlike standard percolation, where failure progresses gradually, interdependent networks can experience abrupt, catastrophic transitions. Notably, networks with heterogeneous degree distributions, typically robust against random failures, become vulnerable in the presence of dependency links, highlighting important design and resilience considerations.

Insights into these phenomena have been obtained through both analytical and numerical studies~\cite{gao2012networks,Havlin2015Mar,Radicchi2015Jul}. In interdependent random networks, the percolation transition is of a hybrid nature: it is discontinuous but also exhibits critical phenomena, such as the presence of critical exponents~\cite{Baxter2012Dec}. In plain percolation, there are two equivalent theoretical descriptions, one based on cluster statistics and the other on the order parameter~\cite{stauffer92}. Yet, these approaches split in the interdependent case, leading to two sets of critical exponents related by scaling relations~\cite{lee16,choi24}. 
In a recent paper, an interesting relationship between the hybrid transition in cascading processes and the spinodal one was shown, leading to the identification of universality classes~\cite{Bonamassa2025Feb}. 

One of the most exciting recent developments is the experimental realization of interdependent percolation in coupled superconducting multilayer systems~\cite{Bonamassa2023Aug}. In this setup, two layers of disordered superconductors are separated by an insulating layer but become effectively coupled through Joule heat. This way, when one superconducting grain transitions to a normal state, it generates heat that causes other grains in the adjacent layer to transition as well. The resulting feedback loop drives a non-local dependency coupling and confirms the theoretical prediction of an abrupt transition. These experimental observations not only validate the theory but also open the door to innovative technological applications based on interdependent dynamics.

\subsection{Spreading processes}
\label{sec:spreading}

In the context of diffusion and spreading dynamics, multilayer networks offer a powerful framework to study how coupled processes may interact, leading to the uncovering of hidden dependencies, feedback loops, and emergent phenomena~\cite{Arruda2018}. This approach is particularly valuable when different types of interactions or dynamical processes influence each other in ways that are not evident in single-layer representations.

\begin{figure*}
\begin{center}
\includegraphics[width=0.6\textwidth]{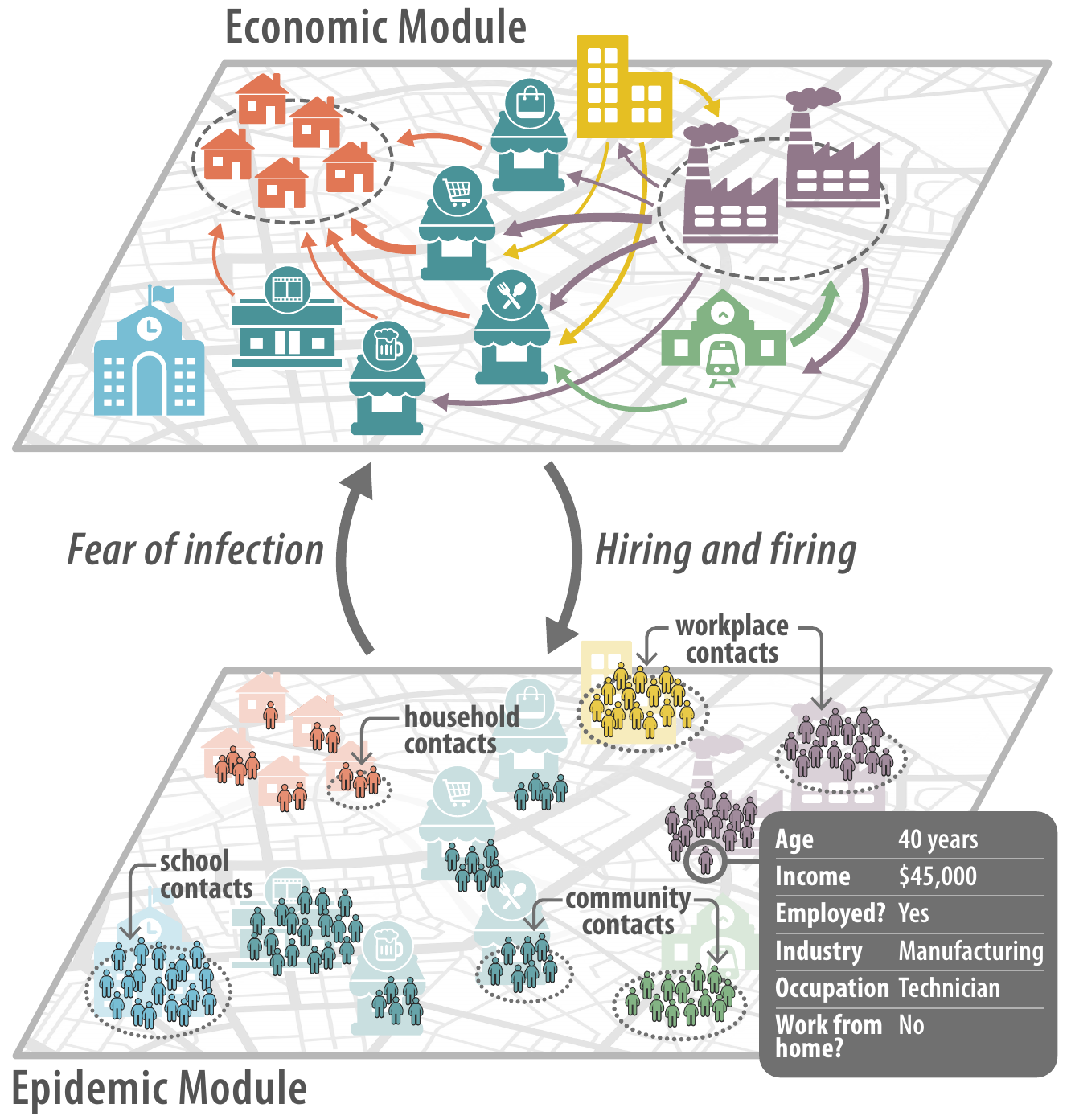}
\caption{Schematic representation of an epidemic-economic model. The figure illustrates the interplay between two very different dynamics. An economic component, which represents the flow of goods and services among industries and to final consumers through an input-output network, and an epidemic component, tracking pathogen exposure in various settings such as workplaces, community/consumption venues, schools, and households via a multilayer contact network. Agents within the system exhibit diverse socio-economic characteristics (see the accompanying box). The tight integration of the economic and epidemic modules is evident: the epidemic module influences economic outcomes by reducing consumption due to infection concerns, while the economic module affects epidemic transmission by modifying workplace and community interactions through industry-specific employment fluctuations.  Reprinted figure with permission from \cite{Pangallo2024Feb}. \copyright 2024 by Springer Nature.}
\label{fig:multi_epid}
\end{center}
\end{figure*}

The range of systems that benefit from this approach is broad. For example, the mutual influence between public opinion and disease spread can be modelled by layering the dynamics of information diffusion with those of epidemic propagation. Such an approach reveals insights into behavioural responses and feedback mechanisms that would otherwise be obscured~\cite{Granell2013Sep,Wang2015Dec,Fosch2023Nov,peng_multilayer_2021}. Multilayer networks also naturally represent systems involving multiple domains with interdependent effects, such as the bidirectional impact between pandemics and economic activity (Fig.~\ref{fig:multi_epid})~\cite{Pangallo2024Feb}. Or systems with a single process but with different spreading properties depending on the type of interaction~\cite{Aleta2020Jul}.

Another compelling example involves the interplay between diseases with distinct transmission mechanisms. Tuberculosis (TB), transmitted via airborne particles, and HIV, spread through sexual contact, represent independent but coexisting threats within a population. Using a multilayer framework, one can explore how interventions or behavioural changes in one disease layer affect the dynamics of the other~\cite{Sanz2014Oct,Bosetti2020Dec,Kinsley2020Sep,Wang2019Sep}. This holistic modelling approach aids in designing integrated public health strategies.

In essence, multilayer networks provide a powerful framework to address the intricate interdependencies between different spreading processes, opening new avenues for understanding, forecasting, and controlling these complex phenomena. 

\subsection{Computational social science}
\label{sec:CSS}

Social systems are inherently characterised by multiple types of relationships that coexist and interact across diverse contexts, timescales, and communication channels. In computational social science, multilayer networks have become a key framework for analysing how behaviours, attitudes, and information propagate through these interconnected pathways. This approach enables the study of complex dynamics such as the interplay between online and offline ties~\cite{hristova2014keep}, cross-platform relationship evolution~\cite{urena2024multiplexity}, and behaviours emerging from interacting relationship types~\cite{garcia2015ideological}.

In political communication, distinct interaction types (support, comments, likes) form separate layers of a multiplex network, each reflecting unique dimensions of polarization and temporal evolution~\cite{garcia2015ideological}. For example, in parliamentary Twitter (now X) networks, echo chambers differ across layers: following networks are highly polarized, while mention networks allow more cross-ideological communication~\cite{del2018echo}. Cross-platform studies find consistent drivers of polarization, such as homophily and selective exposure, even when patterns vary between platforms~\cite{cinelli2021echo}.
In a study of French Twitter users, the multidimensional character of polarization was revealed by unfolding the interaction between the layer of professional politicians and that of ordinary users~\cite{peralta24}.

Temporal multilayer networks reveal that social multiplexity itself is dynamic. Relationship types activate in different contexts, and temporal correlations between layers modify how influence and information spread~\cite{urena2024multiplexity}. Human activity's bursty nature can create both real and spurious correlations~\cite{starnini2017effects}. In signed networks~\cite{diaz2025signed}, research has explored how structural balance and polarization evolve, including dynamical equilibrium where macro-level patterns persist despite micro-level changes~\cite{talaga2023polarization,gonzalez2025evidence}.

In the semantic domain, multilayer networks help uncover how communities structure discourse around complex issues. Modelling words, themes, and temporal evolution as distinct layers reveals patterns of resilience and vulnerability in discussions, such as those following disasters~\cite{randazzo2024multilayer}. Temporal text networks~\cite{vega2018foundations} and multimodal network models~\cite{bonifazi2023multilayer} further support analysis of communication across media and platforms.

Multilayer approaches have also enriched opinion dynamics, where single-layer models often oversimplify how ideas form and spread. 
 This approach provides a powerful framework for exploring how phenomena in one domain can influence behaviour in another. For example,   a model using a communication layer for sharing opinions and an influence layer for observing actions, revealed a range of complex real-world outcomes~\cite{zino_two-layer_2020}. These included rapid, widespread paradigm shifts, the establishment of unpopular norms where innovation adoption fails despite positive sentiment, and persistent community resistance to change.

As discussed in the previous section, the interaction between different processes is a key area of study for multilayer network research. 
A recent work~\cite{amato_opinion_2017} investigated a two-layered network, where agents held opinions on different issues in each layer. While each layer individually tended toward consensus, the interactions between layers could stabilize a mixed state where both opinions coexisted for a long period. This was most pronounced in sparse, positively correlated networks, which facilitated the formation of cross-layer opinion groups. 

The adaptive voter model~\cite{holme06} has been extended to multilayer networks to study how evolving connections affect collective opinion. While multiplexity (the fraction of common nodes across layers) can prevent complete fragmentation, it can also give rise to a new ``shattered'' fragmentation phase~\cite{diakonova_irreducibility_2016}. This phase is characterised by large consensus groups of opposite opinions coexisting with isolated nodes. A similar shattered fragmentation transition in the multiplex adaptive voter model with a notion of triadic closure has been used to describe the empirical distribution of cluster sizes in an online game \cite{klimek2016dynamical}.

Furthermore, a central question in this modelling paradigm is the extent to which such complex multilayer systems can be simplified, especially whether the dynamics on a multilayer network could be reduced to an equivalent single-layer representation. It turns out that conventional aggregation procedures fail to capture key non-linear effects, such as the prolonged lifetime of multiplex states at low interlayer connectivity~\cite{diakonova_irreducibility_2016}. This work suggests that there are fundamental, non-reducible differences between multilayer dynamics and their single-layer approximations, highlighting the necessity of explicitly modelling these intricate, interconnected structures to fully understand complex social phenomena.

Altogether, the integration of multilayer networks into computational social science enables more realistic, nuanced models of human behaviour. As interactions increasingly span digital and physical spaces, these tools are essential for understanding polarization, social cohesion, and collective action in a complex, connected world.

\subsection{Economic and financial systems}
\label{sec:economics}

Economics and finance are inherently based on networked systems. They describe interactions among diverse actors, including individuals, firms, governments, and institutions, that trade, collaborate, invest in, or oversee one another. Markets can be seen as platforms that promote the constant evolution and reshaping of existing networks among different actors. For instance, financial markets facilitate the creation and deletion of links between investors and firms, while the labour market connects workers with employers. New technological platforms can also rapidly reshape existing networks, as in the case of Uber transforming urban transportation labour markets. In all these cases, economic and financial systems heavily rely on information flowing between entities belonging to different networks, such as price signals, market trends, political news, and policy decisions.

Based on this premise, it is not surprising that multiplex and multilayer networks may occupy a central position in the analysis and understanding of economic and financial systems. While challenges such as data access and system complexity have limited the full adoption of these approaches, notable studies already highlight their potential. One example is the use of a multiplex global financial network, in which nodes represent countries, edges encode cross-country financial assets of various types, and layers represent asset types. The number of systemically important countries almost doubles when the heterogeneity of financial exposures is considered (i.e., when using a multiplex network), compared to results from an associated aggregate global financial network (i.e., a single-layer network), long regarded as the standard model for such investigations~\cite{delriochanona2020multiplex}.

Another study constructed a multiplex network of financial institutions, with layers corresponding to different classes of derivative contracts (e.g., interest rate, credit, foreign exchange), revealing how liquidity shocks propagate across markets~\cite{bardoscia2019multiplex}. A linked multiplex network incorporating ownership, shared board members, R\&D partnerships, and stock correlations among companies in the UK and Germany uncovered a higher level of corporate interconnectedness than previously reported~\cite{jeude2019multilayer}.

Additional insights have come from analysing the multilayer structure of international trade, where nodes represent countries, layers correspond to industries, and links connect buyers and sellers. A multilayer nestedness metric captured structural organisation more effectively than traditional bipartite models~\cite{alves2019nested}. Similarly, the integration of ownership, innovation collaborations, and board interlocks has shed light on the formation of innovation clusters and their influence on market concentration~\cite{glattfelder2013decoding}.

Systemic risk, a core concern in financial stability, is deeply rooted in network interdependencies. Multilayer models have advanced our understanding of how interconnections amplify or mitigate risk~\cite{battiston2012debtrank,caccioli2014stability}. Studies of global supply chains have also benefited from multilayer perspectives, identifying fragile sectors such as pharmaceuticals, food, and services as critical nodes in system-wide vulnerability~\cite{gomez2020fragility}. Lastly, the cryptocurrency ecosystem—comprising thousands of token layers linked by users, exchanges, and dark markets—presents a complex but largely unexplored multilayered network.

Altogether, these examples underscore the promise of multilayer network analysis in uncovering hidden interdependencies, enhancing systemic risk models, and informing economic policy in an increasingly interconnected world.

\subsection{Infrastructure networks}
\label{sec:infra}

Infrastructure systems, such as power grids, transportation networks, and communication systems, are important examples where multilayer networks provide insights into interdependencies and weaknesses. Among these, power grids received significant attention in the early days of the field, particularly after it was shown that interdependencies can make networks more fragile~\cite{buldyrev_2010} (see also Section~\ref{sec:failures}.) This work was motivated by the 2003 blackout in Italy, where failures in power stations led to disruptions in communication networks, which in turn caused further failures in the power grid itself. Subsequent studies confirmed these findings and investigated how various types of interconnections influence robustness~\cite{gao2012networks}.

This research line has since diversified, exploring network fragility from various angles~\cite{Duan2019Nov,artime2024robustness}. For example, some studies have examined cascading failures using load-sharing dynamics~\cite{Brummitt2012Mar,Artime2020Sep}, while others have focused on designing inherently more robust networks~\cite{Reis2014Oct,Latora2005Jan}. There has also been growing interest in identifying early warning signs of collapse, a challenge relevant not only to infrastructure, but also to biology, economics, and ecology~\cite{Duan2019Nov}.

More recently, researchers have noted that traditional interdependent network models often assume "hard failures," where the collapse of a node in one layer immediately causes failure in a dependent node. Yet, such failures are rare in practice. Instead, attention has shifted toward modelling interdependencies during recovery processes, where one network relies on resources from another to regain functionality~\cite{Danziger2022Feb}.

\begin{figure*}
\begin{center}
\includegraphics[width=\textwidth]{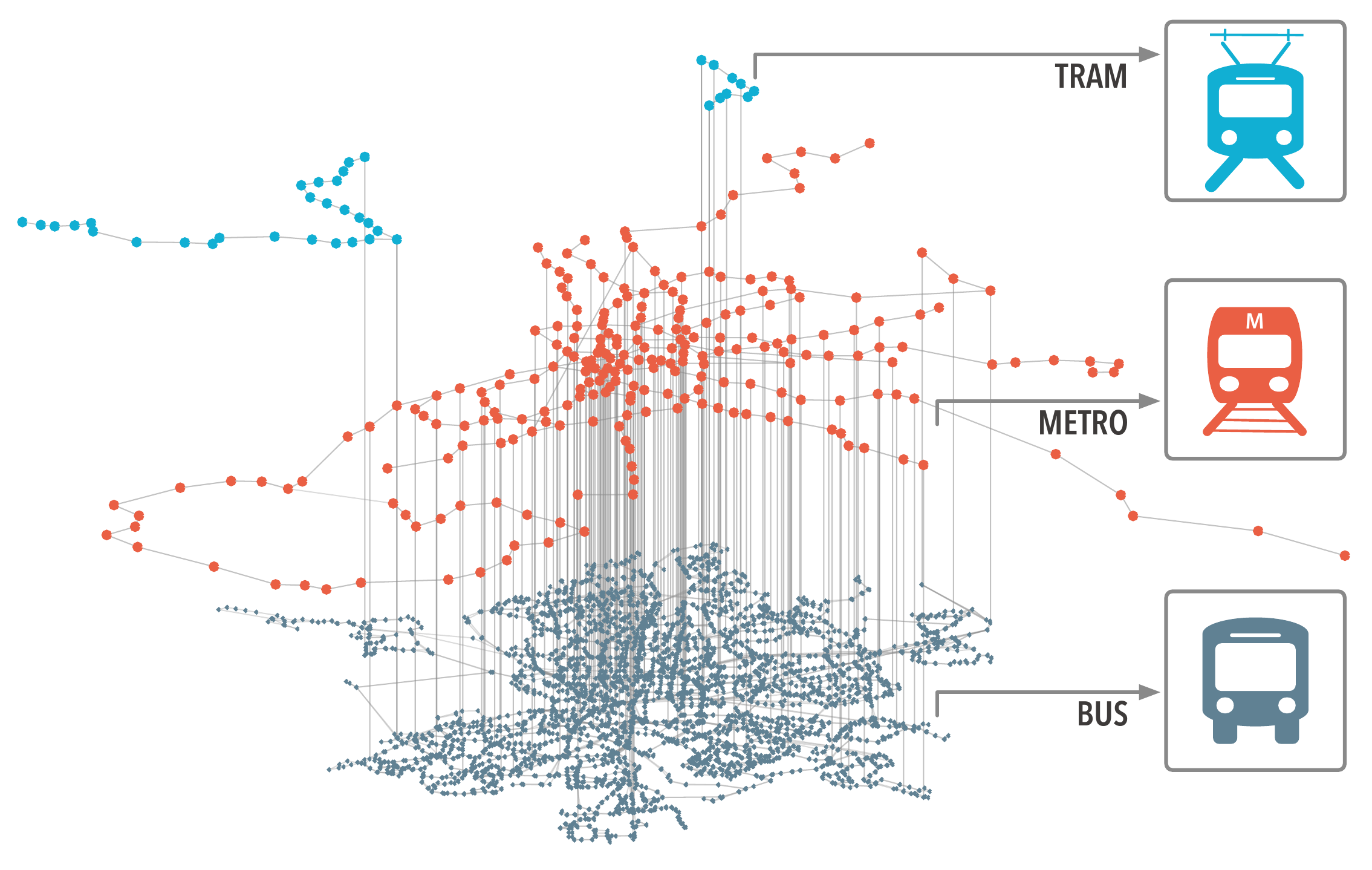}
\caption{Multilayer representation of Madrid's public transportation system in 2016. The network is organized into three layers, each representing a different mode of transportation: the top layer (cyan nodes) corresponds to the tram system, the middle layer (orange nodes) to the metro system, and the bottom layer (dark blue nodes) to the bus system. Vertical links connect stops of different modes located within a 150-meter radius, illustrating intermodal connections. Adapted figure with permission from~\cite{Aleta2019Mar}. \copyright 2019 by Annual Reviews.}
\label{fig:transport_Madrid}
\end{center}
\end{figure*}

Transportation systems, much like power grids, have been extensively studied through the lens of multilayer networks~\cite{Alessandretti2022Jul}. Early work modelled the European air transportation network to assess the impact of flight cancellations on network performance~\cite{Cardillo2013Jan}. Urban mobility networks further highlight the complexity of multilayer infrastructure in metropolitan areas. Cities rely on multimodal transportation—streets, subways, buses—that together form the backbone of mobility (Fig.~\ref{fig:transport_Madrid})~\cite{Aleta2017Mar,Natera2020Jun}. Studies have shown that poor synchronisation between modes can lead to significant inefficiencies, including delays and reduced accessibility~\cite{Gallotti2014Nov,Gallotti2015Jan}. Moreover, navigating these systems can challenge human cognitive limits, making trip planning difficult in highly interconnected urban environments~\cite{Gallotti2016Feb}.

From a dynamical standpoint, several researchers have explored congestion in multilayer transportation systems~\cite{Du2016Jan,Lampo2021Mar}. Notably, increasing speed in one layer—such as subways—can sometimes produce counter-intuitive effects, like congestion at terminal stations due to imbalanced flow dynamics~\cite{Strano2015Oct}. Expanding the scope beyond transportation modes, urban mobility was analysed by categorizing daily activities (e.g., visiting tourist attractions or sports venues) into distinct layers, offering a multidimensional representation of urban systems~\cite{Gallotti2021Dec}.

In summary, multilayer network analysis has deepened our understanding of infrastructure systems, from cascading failures in power grids to mobility challenges in urban transport. These insights help identify vulnerabilities, design more resilient systems, and optimise performance, ultimately enhancing the functionality and robustness of essential infrastructure in modern society.

\subsection{Ecological networks}
\label{sec:ecol}

Traditionally, ecological networks have been represented as single-layer networks, focusing on single types of interactions, such as predator-prey or plant-pollinator interactions~\cite{Tylianakis2017Nov}. Similarly, although biological communities are intrinsically dynamic, with species and interactions changing over time, temporal dynamics have often been analysed using independent snapshot networks. Thus, ecological systems provide a perfect example of where multilayer networks can play a crucial role by allowing us to simultaneously capture different types of pairwise interactions~\cite{Garcia-Callejas2018Jan} or temporal and spatial dimensions, providing a more complete view of ecosystem functioning~\cite{Hutchinson2019Feb,Finn2019Mar}.

One study used a five-year dataset on bird-seed dispersal interactions to analyse species turnover and persistence, identifying a stable core of species that consistently supported network structure over time~\cite{Costa2020Sep}. Similar insights have emerged from studies of plant-pollinator networks that tracked the spatial and temporal persistence of species and interactions~\cite{Hervias-Parejo2023Aug}, and from work emphasizing the advantages of multilayer approaches over traditional habitat discretisation methods~\cite{Timoteo2018Jan}.

The common binary classification of species as mutualists or antagonists has also been challenged using multilayer frameworks. In one example, seed viability from faecal samples of wild-trapped mammals showed that some species acted as both antagonists and mutualists for the same plant species, depending on context—revealing a much richer set of interactions than standard models capture~\cite{Genrich2017Mar}. Multilayer analysis of frugivory and nectarivory interactions has further revealed non-obvious rules of community assembly~\cite{Mello2019Nov}, while quantifying link strength in networks of pollinators and seed dispersers uncovered that only a subset of individuals truly connects both ecological functions~\cite{Hervias-Parejo2020Nov}.

Animal social networks also benefit from multilayer analysis, as interactions often span multiple social contexts. A study of baboons with layers representing grooming and association interactions demonstrated that an individual's centrality ranking can vary dramatically depending on which layer or aggregation is considered~\cite{Finn2019Mar}. This underlines the importance of selecting relevant interactions when drawing ecological or behavioural inferences.

Recently, a global multilayer modelling framework spanning six ecological functions—pollination, herbivory, seed dispersal, decomposition, nutrient uptake, and fungal pathogenicity—enabled the identification of keystone species, defined by their critical role in maintaining ecosystem function regardless of abundance~\cite{Hervias-Parejo2024Oct}.

Beyond species interactions, multilayer networks have been applied to physical ecological systems. For instance, river deltas have been modelled as multilayer transport networks~\cite{Tejedor2018Sep}, and ecosystem restoration has been studied by tracking structural and functional changes in multilayer networks over time~\cite{Moreno-Mateos2020May}. Comparing evolving network states to reference conditions helps assess recovery trajectories.

Despite the promise of multilayer networks in ecology, challenges remain. These include the difficulty of collecting fine-grained interaction data~\cite{Seibold2018Oct}, the complexity of comparing fundamentally different interaction types~\cite{Hervias-Parejo2024Oct}, and the ongoing need to develop multilayer metrics tailored for ecological applications~\cite{Garrido2023Aug}. Addressing these limitations will be key to further advancing ecological insights through multilayer network approaches.

\subsection{Science of science}
\label{sec:scisci}

The science of science is the study of the structure and evolution of science via the analysis of data on scientists and their interactions~\cite{fortunato18,wang21}. Although the field dates back several decades~\cite{price63}, it has grown rapidly in recent years due to the increasing availability of data on scientific productivity, collaboration, citation, mobility, funding, etc., and to advances in analytical tools, especially in machine learning. Network science has heavily contributed to the development of the field, due to the many networks that can be built from the data. 

Multilayer networks have brought added value to this domain by capturing different dimensions of scientific interactions. Early studies showed that combining collaboration networks across journals or merging collaboration and citation networks into a single layer can obscure meaningful correlations~\cite{menichetti2014weighted}. A strong correlation between community structures was observed across layers of a multiplex collaboration network of physicists, where each layer represents a different subfield~\cite{battiston16}.

The multilayer framework has also revealed intricate dependencies between collaboration and citation behaviours. For example, papers authored by scientists with a large number of previous collaborators tend to gain citations more quickly, but also fade faster in attention~\cite{zingg20}. Such dependencies can be exploited to develop more accurate generative models of citation networks~\cite{nanumyan20}. Moreover, multilayer networks are better than single-layer ones when it comes to predicting future links, specifically future collaborations between scientists. For this task, besides the collaboration network itself, useful layers may consist of citations between scientists and semantic closeness of the keywords of their papers~\cite{tuninetti21}. Other relevant network types include co-citation and co-venue networks, where two scientists are connected if they cite the same papers or publish in the same journals~\cite{pujari15}. 

Ranking scientists by impact is another area enriched by multilayer methods. Traditional indicators such as raw citation counts or the h-index~\cite{hirsch05} may overlook important structural information. Multilayer networks make it possible to compute composite impact scores that account for multiple dimensions, such as citations and collaborations, either by aggregating layer-specific rankings~\cite{pradhan17} or through diffusion-based algorithms like multilayer PageRank~\cite{omodei17}. The framework also extends to innovation studies, particularly in analysing patent citation networks. For example, a multiplex network in which layers correspond to different patent-granting offices has been shown to better capture the technological classification of patents than aggregated representations~\cite{higham22}.

Overall, multilayer networks provide a richer, more nuanced view of scientific ecosystems, enhancing our ability to model, predict, and evaluate scientific activity across various dimensions.

\subsection{Climate science}
\label{sec:climate}

Network analysis has become a valuable complement to numerical modelling in climate science~\cite{ludescher21}. In climate networks, nodes typically represent geographic locations, and links indicate similarities between these locations based on relevant physical quantities such as temperature, precipitation, atmospheric pressure, or CO$_2$ concentration. These similarities are quantified using various linear and non-linear measures computed on time series, including Pearson correlation, mutual information, event synchronisation, transfer entropy, partial correlation, or Granger causality.

Most early studies focused on single-variable networks, but recent work has begun to leverage multilayer networks to capture interdependencies among different climate variables or the same variable under different conditions. For instance, two-layer networks constructed from atmospheric pressure fields at different altitudes have revealed significant insights into the general circulation and stratification of the atmosphere~\cite{donges11}.

Multilayer networks also offer powerful tools for assessing climate vulnerability. By integrating climatic and non-climatic factors across multiple layers, researchers can analyse how exposure, sensitivity, and adaptive capacity interact to shape a community's vulnerability to climate change~\cite{debortoli18}. Similarly, multilayer behavioural networks that integrate supply-side production and demand-side consumption decisions have been used to explore how climate shocks propagate through social and economic systems~\cite{naqvi21}.

Air pollution is another area where multilayer networks have yielded novel insights. A multilayer network study linking the 500-hPa pressure height (geo-potential height) with surface-level air pollution in China and the United States demonstrated how Rossby waves (large-scale atmospheric waves induced by Earth's rotation) influence pollution through their effects on cyclone and anticyclone systems~\cite{zhang19}.

The study of CO$_2$ dynamics, a major driver of global warming, has also benefited from multilayer approaches. One interlayer climate network used atmospheric mid-tropospheric CO$_2$ records and ground-level surface air temperature (SAT) data to investigate interactions between CO$_2$ and SAT~\cite{ying20}. This analysis revealed strong interlayer connections, with the most significant links often spanning more than 7,000 km, a distance consistent with the wavelength of atmospheric Rossby waves. A follow-up study removed the influence of Rossby waves to isolate regions where CO$_2$ concentrations are most directly related to SAT variations~\cite{ying21}.

Thus, multilayer network analysis enables a more holistic understanding of the Earth's climate system by revealing complex interdependencies among climatic variables and between environmental and societal factors. These insights can inform more effective strategies for climate adaptation, risk assessment, and policy development.

\subsection{Network medicine}
\label{sec:netw_medicine}

Network medicine applies network science to biological and medical data to better understand, prevent, and treat diseases~\cite{barabasi11}. Biological systems encompass a wide variety of networks, including protein–protein interactions (PPIs), metabolic networks, gene regulatory networks, or cell–cell communication networks. Traditionally analysed in isolation, recent years have seen a shift toward integrating diverse omics data (genomics, proteomics, metabolomics, metagenomics, phenomics, and transcriptomics), leveraging multilayer networks to provide a more holistic view of complex biological processes.

Multi-omics integration, especially when coupled with longitudinal data, enables researchers to observe dynamic interactions between biological layers, identify key components of system development, and better understand complex phenotypes. Among biological networks, PPIs are some of the most extensively studied. In multiplex PPI networks, layers may represent interactions in different types of cells (e.g., normal vs. cancerous), different interaction types (e.g., genetic vs. physical), or different data sources (e.g., co-expression, co-annotation)\cite{rai17,yu20,zhao16}. Such representations have been used to detect biologically enriched modules\cite{mangioni18} and to improve the prediction of protein functions.

\begin{figure*}
\begin{center}
\includegraphics[width=0.7\textwidth]{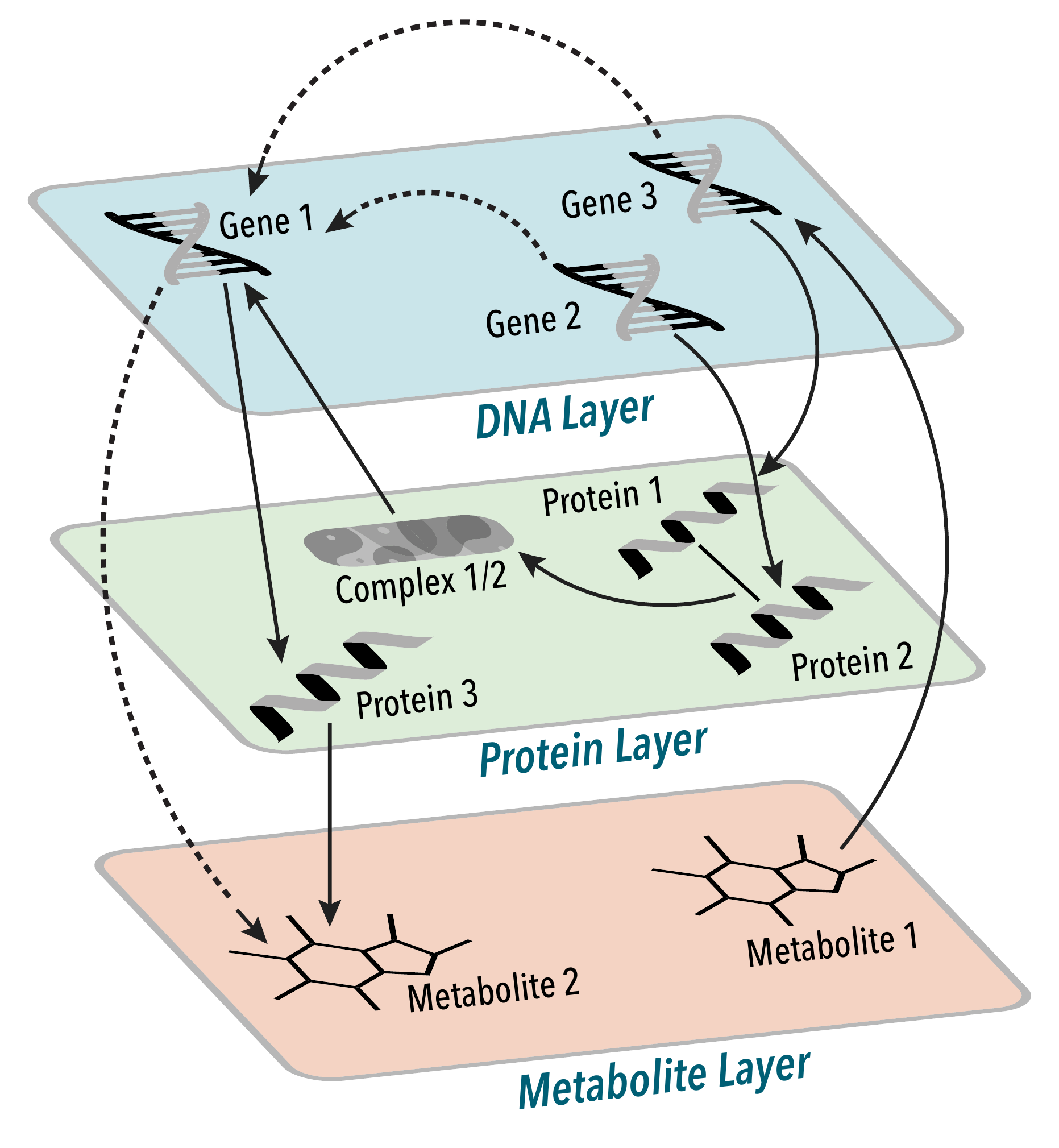}
\caption{Multilayer network featuring different omics (genes, proteins, metabolites) and their mutual interactions/associations. Adapted figure with permission from~\cite{hammoud20}. \copyright 2020 by Springer Nature.}
\label{fig:multilayer_med}
\end{center}
\end{figure*}

More frequently, scholars build multilayer networks by combining different omics, as illustrated in Fig.\ref{fig:multilayer_med}. In one of the first works, layers consisted of transcription factor co-targeting, microRNA co-targeting, protein-protein interaction, and gene co-expression networks. Communities of genes were extracted by combining partitions found in the individual layers via consensus clustering~\cite{lancichinetti12}. Enrichment analysis on the genes belonging to the same community revealed a set of candidate cancer drivers~\cite{cantini15}. 

Robustness is a prominent feature of most biological systems. By studying the spreading of perturbations in a multilayer network composed of a gene regulatory layer, a protein–protein interaction layer, and a metabolic layer,  one can characterise the contribution of each gene to the overall system’s robustness:  the most influential genes are enriched in essential and cancer genes~\cite{liu20}. An analysis of the spreading of perturbations through the three layers of a graph representation of the integrative system of E. coli’s gene regulation and metabolism shows that small perturbations, originating in the gene regulatory domain, typically trigger far-reaching system-wide cascades, while small perturbations in the metabolic domain tend to remain more local and trigger much smaller cascades~\cite{klosik17}. 

Random walk dynamics offer a powerful framework for analysing multilayer biological systems. For example, random walks across networks of disease-perturbed proteins, drug targets, and biological functions help trace how drug effects propagate~\cite{ruiz21}. Other studies use random walks on multi-omics networks (combining layers from mRNA expression, PPIs, KEGG pathways, and Gene Ontology terms) to map omics molecules to biological functions~\cite{bodein22}. This methodology has also been applied to COVID-19 datasets, enabling insight into disease severity and therapeutic targets~\cite{agamah24}. Single-cell multi-omics multilayer networks have further improved predictions of transcription factor binding sites, regulatory regions, and gene communities~\cite{trimbour24}.

Another area of interest involves disease–disease networks, where nodes are diseases and edges represent functional, semantic, or co-morbidity associations. For instance, two-layer networks combining gene-based and ontology-based similarities have been used to identify diseases with potential shared drug treatments~\cite{cheng14}. Multiplex networks with phenotypic, genetic, pathway-based, and toxicogenomic co-morbidities have revealed the dominant causative mechanisms for hundreds of diseases~\cite{klimek16}. In another study, a bilayer network of 779 diseases—connected via shared genes and symptoms—enabled the detection of disease communities using multiplex Infomap~\cite{halu19,dedomenico2015identifying}. Similarly, during the COVID-19 pandemic, multilayer approaches were instrumental in understanding the disease. The CovMulNet19 network integrated SARS-CoV-2 proteins, host proteins, related diseases, symptoms, and drug targets, and its analysis revealed high similarity between COVID-19 and diseases that are also known patient risk factors~\cite{verstraete20}.

Finally, the growing field of network embeddings has facilitated the integration of heterogeneous biological datasets. By embedding multilayer biological networks into geometric spaces, researchers have improved tasks like drug–target interaction prediction and biomarker identification~\cite{zitnik17,jagtap22} (see Section~\ref{sec:ML}).

In short, multilayer networks are proving essential for capturing the complexity of biological systems and improving our understanding of disease mechanisms, interactions, and potential treatments.

\subsection{Network neuroscience}
\label{sec:netw_neuro}

Neuroscience is fundamentally centred on the study of the brain as a vast, interconnected system of neurons and brain regions. This field encompasses not only the structural and functional organisation of the brain but also the dynamic processes that underpin cognition, behaviour, and neurological health. Networks provide a natural framework for capturing these complexities by representing the brain as a system of nodes and edges. Over the past decade, network neuroscience \cite{bassett2017network,barabasi2023neuroscience} has emerged as one of the most exciting and influential fields, offering new insights into the brain’s organisation and its intricate functions.
 
Brain networks \cite{bullmore2009complex} are mathematically modelled systems where nodes represent either neurons or specific brain regions, while edges capture various forms of connectivity. These connections may represent structural connectivity, which reflects the physical pathways linking brain regions and is often derived from diffusion MRI or diffusion tensor imaging. Alternatively, they may capture functional connectivity, which describes statistical relationships in neural activity between regions, as observed through functional MRI, electroencephalography, or magnetoencephalography. A third type, effective connectivity, goes a step further by modelling causal interactions and directional relationships between regions, often using techniques like Granger causality or dynamic causal modelling \cite{nicosia2013}. Together, these distinct forms of connectivity provide a rich and multifaceted view of the brain’s organisation and activity.
 
Building on this foundation, multilayer networks offer a sophisticated extension for exploring brain connectivity and dynamics \cite{de2017multilayer,presigny22}. These models expand the traditional network framework by incorporating multiple interconnected layers of nodes and edges, each representing a different dynamic or scale of analysis \cite{betzel2017multi}. Within an individual, for example, these layers can correspond to distinct types of connectivity, such as structural and functional networks, with interlayer links capturing relationships between them \cite{vaiana2018multilayer}. Similarly, temporal transitions in dynamic functional connectivity can be represented as changes within and across layers or states \cite{sporns2021dynamic}. Beyond individuals, multilayer networks also enable the comparison of connectivity patterns across populations by modelling interlayer relationships that represent similarities between individuals or groups \cite{wilson2020analysis,simas2015}. This versatility allows multilayer networks to bridge gaps between data types and scales, revealing insights that would be inaccessible using single-layer approaches \cite{battiston18, mandke2018comparing,presigny24}.
 
The ability to integrate and analyse these diverse sources of data has led to important studies in areas such as human learning \cite{bassett2017learning}, individual differences \cite{betzel2019community}, the diagnosis and understanding of neurological and mental disorders \cite{stam2014modern}, among others like the relation between social environment and variation in brain network organisation \cite{merritt2024stability}. 
 
Additionally, even though traditional network models focus mainly on pairwise relationships, higher-order interactions can be pulled into the multilayer network paradigm. For instance, one could consider multiple layers describing different orders of interaction, an approach that has contributed to new advances in the understanding of human brain evolution and cognition~\cite{luppi2022synergistic}, and the relation between brain activity and behaviour~\cite{santoro2024higher}. In addition, higher-order structures can help uncover the relationship between structure and function in the brain~\cite{battiston17}.
 
In summary, the multilayer networks framework represents a frontier in network neuroscience, enabling researchers to integrate diverse data types and explore the brain’s multidimensional nature. Continued methodological advances will undoubtedly unlock clearer insights into brain function and dysfunction, fostering progress in both neuroscience and clinical applications.

\section{Data \& Tools\label{sec:data}}

Given the growing interest in network science, a wide array of resources has become available. Yet, despite substantial progress, challenges remain in terms of data availability, standardisation, and tool interoperability. This section highlights the most relevant public data repositories and software libraries that support advanced network analysis and visualisation.

\subsection{Data}

Network data comes from a wide range of disciplines such as biology, sociology, economics, neuroscience, and more. While numerous repositories exist for general-purpose network datasets, multilayer and temporal network datasets are comparatively less common and often tailored to specific domains. Furthermore, the lack of standardised formats can hinder reproducibility and comparative studies across methods.

Nevertheless, various curated and open-access repositories offer high-quality datasets that have become benchmarks in the field. Table~\ref{tab:network_data_sources} lists key sources of network datasets across domains. It is worth noting that the table only includes major and widely used repositories. Many individual researchers also make their datasets publicly available on personal or project websites. While this is an invaluable contribution to the community, it also highlights the fragmentation of the field and the lack of consistent standards for sharing and documenting network data, although there is a lot of effort to improve this situation.

\begin{table}[H]
\centering
\begin{tabular}{lrl}
\toprule
\textbf{Repository} & Networks & \textbf{Reference} \\
\hline
\texttt{Netzschleuder} & 163735 & \cite{Netzschleuder}\\
\texttt{Network Repository} & 5664 & \cite{nr} \\
\texttt{KONECT} & 1326 & \cite{konect} \\
\texttt{ICON} & 704 & \cite{icon} \\
\texttt{SNAP} & 127 & \cite{snapnets}\\
\texttt{UCI} & 35 & \cite{DuBois:2008}\\
\hline
\end{tabular}
\caption{Examples of publicly available network datasets, sorted by number of networks available as of September 2025. Some repositories include multilayer, temporal, and higher-order networks.}
\label{tab:network_data_sources}
\end{table}

While many general-purpose datasets are readily accessible and well-documented, multilayer and higher-order data remain underrepresented. A key issue is the lack of standardised file formats to represent these complex structures, which can complicate interoperability between tools and comparisons across studies. 

\subsection{Software Tools}

The development of dedicated software libraries has significantly accelerated the field of network science. A wide variety of tools exist for analysis, simulation, and visualisation of networks. Table~\ref{tab:network_libraries} summarizes major software libraries, including their main focus and programming language. It is important to note that this overview focuses on general-purpose tools for the structural analysis and visualisation of complex networks. Tools primarily focused on specific functional processes or computational workflows, such as those for graph neural networks~\cite{pytorch} or epidemic simulations~\cite{czuba2024networkdiffusion,gozzi2025,Miller2019-ty}, are not included.

However, no single tool dominates the field. Even for general-purpose networks, researchers choose between different libraries based on project requirements. For example, while \texttt{networkX} is widely used for its ease of use and versatility, alternatives such as \texttt{igraph} or \texttt{graph-tool} offer superior performance on large networks due to optimised C/C++ backends. Thus, selecting the appropriate tool often involves trade-offs between ease of use, scalability, feature set, and visualisation capabilities. 

\begin{table*}
\centering
\begin{tabular}{llllc}
\hline\hline
\textbf{Type}                             & \textbf{Name}                                  & \textbf{Main Focus}            & \textbf{Language} & \textbf{Reference}                                \\ 
\hline
\multirow{20}{*}{\rotcell{Analysis}}      & \texttt{DyNetX}               & Dynamic networks               & Python            & \cite{dynetx}                    \\
                                          & \texttt{HAT}                  & Higher-order networks          & Python, MATLAB    & \cite{Pickard2023Jun}            \\
                                          & \texttt{graph-tool}           & General-purpose networks       & Python            & \cite{peixoto_graphtool_2014}  \\
                                          & \texttt{HGX}                  & Higher-order networks          & Python            & \cite{lotito2023hypergraphx}     \\
                                          & \texttt{HyperNetX}            & Higher-order networks          & Python            & \cite{hypernetx}                 \\
                                          & \texttt{iGraph}               & General-purpose networks       & Python, R, C      & \cite{igraph}                    \\
                                          & \texttt{MultilayerGraphs.jl}  & Multilayer networks             & Julia             & \cite{MultilayerGraphs}          \\
                                          & \texttt{multinet}             & Multilayer networks             & R                 & \cite{multinetR}                 \\
                                          & \texttt{NetworKit}            & Large-scale networks           & Python            & \cite{networkit}                 \\
                                          & \texttt{networkX}             & General-purpose networks       & Python            & \cite{networkx}                  \\
                                          & \texttt{NDlib}                & General-purpose networks       & Python            & \cite{Rossetti2018Feb}           \\
                                          & \texttt{pathpy}               & Higher-order temporal networks & Python            & \cite{Hackl2021Apr}              \\
                                          & \texttt{pymnet}               & Multilayer networks             & Python            & \cite{pymnet}                    \\
                                          & \texttt{pyunicorn}            & General-purpose networks       & Python            & \cite{Donges2015Nov}             \\
                                          & \texttt{Reticula}             & Higher-order temporal networks & Python            & \cite{badie2023reticula}         \\
                                          & \texttt{SimpleHypergraphs.jl} & Higher-order networks          & Julia             & \cite{Spagnuolo2020Mar}          \\
                                          & \texttt{straph}               & Temporal networks              & Python            & \cite{Latapy2018Dec}             \\
                                          & \texttt{tacoma}               & Temporal networks              & Julia             & \cite{tacoma}                    \\
                                          & \texttt{TGLib}                & Temporal networks              & Python, C++       & \cite{Oettershagen}              \\
                                          & \texttt{tnetwork}             & Temporal networks              & Python            & \cite{tnetwork}                  \\
                                          & \texttt{XGI}                  & Higher-order networks          & Python            & \cite{Landry_XGI_2023}         \\ 
\hline
\multirow{14}{*}{\rotcell{Visualisation}} & \texttt{Cytoscape}            & General-purpose networks       & GUI               & \cite{shannon2003cytoscape}      \\
                                          & \texttt{Gephi}                & General-purpose networks       & GUI               & \cite{bastian2009gephi}          \\
                                          & \texttt{ggraph}               & General-purpose networks       & R                 & \cite{ggraph}                    \\
                                          & \texttt{Graphia}              & General-purpose networks       & GUI               & \cite{Freeman2022Jul}            \\
                                          & \texttt{MultiNet}             & Multivariate networks          & Web-based         & \cite{2023_nsf_multinet}       \\
                                          & \texttt{multiNetX}            & Multilayer networks            & Python            & \cite{multinetX}                 \\
                                          & \texttt{muxViz}               & Multilayer networks            & R, Web-based      & \cite{muxviz}                    \\
                                          & \texttt{Netminer}             & General-purpose networks       & GUI               & \cite{Ghim2014}                  \\
                                          & \texttt{Netwulf}              & General-purpose networks       & Python, Web-based & \cite{Aslak2019Oct}              \\
                                          & \texttt{NodeXL}               & General-purpose networks       & Excel             & \cite{nodexl}                    \\
                                          & \texttt{Pajek}                & General-purpose networks       & GUI               & \cite{Batagelj2004}              \\
                                          & \texttt{py3Plex}              & Multilayer networks             & Python            & \cite{Skrlj2019}                 \\
                                          & \texttt{Sigma.js}             & General-purpose networks       & JavaScript        & \cite{sigmajs}                   \\
                                          & \texttt{Tulip}                & General-purpose networks       & GUI               & \cite{Auber2012Jan}              \\
\hline\hline
\end{tabular}
\caption{Libraries and tools for network science analysis and visualisation, with support for multilayer, temporal, and higher-order networks. ``Main focus'' refers to the area where the software is primarily applied; however, some tools can support multiple network types.}
\label{tab:network_libraries}
\end{table*}

\section{Outlook}
\label{sec:outlook}

The past decade has shown that multilayer networks are not merely a convenient generalization of graph theory but a fundamentally new modelling language. The field has now reached a stage where the central question is no longer how to adapt single-layer tools, but how to identify the truly multilayer mechanisms that cannot be reduced away. Several theoretical directions emerge from this perspective. One concerns the search for minimal descriptions: determining when layers encode irreducible information and when they can be compressed. Another direction concerns the interplay between multilayer topology and cross-layer dynamical coupling, where instabilities, hybrid phase transitions, metastability, or coexistence of dynamical regimes can arise due to the multiplex nature of the system. Understanding the general principles that govern these phenomena remains a major open challenge. At the same time, the field’s inherently multidisciplinary nature complicates the picture: multilayer approaches are used in physics, biology, economics, computer science, and the social sciences, often with different assumptions, terminologies, and modelling priorities. This diversity makes communication across communities difficult, yet it also underscores the value of multilayer networks as a common representational language capable of unifying disparate viewpoints. 

At the methodological frontier, several additional opportunities demand attention. The first involves integrating multilayer representations with temporal and higher-order structures. Many real systems combine heterogeneous interaction types with time-varying group interactions, but current modelling frameworks still tend to treat these features separately. Developing unified theoretical tools capable of capturing polyadic interactions, memory effects, and cross-layer dependencies simultaneously is an important next step. A second direction concerns inference, where the rise of generative models, statistical reconstruction techniques, and machine-learning-based embeddings is beginning to provide new ways to infer hidden layers and mesoscale structures. Yet, understanding the limits of inference remains unresolved. A third challenge lies in controllability and optimisation: while single-layer structural controllability is now well understood, its multilayer analogue is largely unexplored, especially when layers interact through non-linear or time-dependent couplings. Finally, the rapidly growing integration between network science and artificial intelligence opens a promising but underdeveloped path. Deep learning architectures are themselves multilayered systems whose internal dynamics could benefit from tools developed in multilayer network science, potentially contributing to interpretability and robustness in AI.

Across domains, applications of multilayer network science are entering a phase in which data and theoretical tools are finally becoming aligned. A key opportunity lies in modelling coupled processes in the wild: infrastructure failures intertwined with cyber systems, economic–epidemic feedbacks, multi-omics interactions, or the joint evolution of online and offline social behaviour. These problems require models that simultaneously encode structural heterogeneity, temporal evolution, and cross-layer feedback, which are precisely the strengths of the multilayer paradigm. A central challenge, however, is that multilayer data and tools originate from many different fields, each with its own conventions, formats, and epistemic goals. This heterogeneity makes it difficult to compare studies or integrate datasets, yet it also reflects the breadth of the systems that multilayer network science can represent. As common standards and interoperable software mature, they provide a platform for genuine cross-disciplinary exchange, helping the field not only accumulate data but also communicate across communities. In scientific domains ranging from ecology to neuroscience, such integrative approaches are beginning to reveal how processes at different scales coordinate, interfere, or reinforce each other.

Looking ahead, progress will depend critically on the availability of high-resolution multilayer data, standardised formats, and interoperable tools, but also on conceptual clarity about what it means for a system to be truly multilayer. If these developments continue in parallel, multilayer network science is poised to move from descriptive analyses to predictive, actionable models capable of guiding interventions, optimising coupled infrastructures, and informing policy in domains where interdependence is no longer optional to ignore. As the complexity of modern systems continues to grow, the multilayer perspective will remain essential for understanding the intertwined processes that characterise the natural and engineered world.

\begin{acknowledgements}
    All authors acknowledge the support of the AccelNet-MultiNet program, a project of the National Science Foundation. We thank M. Clarin from COSNET Lab-BIFI and N. Samay from NetSI for helping with the figures. Funding support for this article was provided by the National Science Foundation (Award \#1927425 and \#1927418). A.S.T. acknowledges support by FCT – Fundação para a Ciência e Tecnologia – through the LASIGE Research Unit, ref. UID/408/2025. A.A. acknowledges support from the grant RYC2021‐033226‐I funded by MCIN/AEI/10.13039/501100011033 and the European Union NextGenerationEU/PRTR. A.A. and Y.M. were partially supported by the Government of Aragon, Spain, and ERDF "A way of making Europe" through grant E36-23R (FENOL), and by Grant No. PID2023-149409NB-I00 from Ministerio de Ciencia, Innovación y Universidades, Agencia Española de Investigación (MICIU/AEI/10.13039/501100011033) and ERDF "A way of making Europe". G.F.A was partially supported by the São Paulo Research Foundation (FAPESP), through grants 2024/16711-8 and 2025/04409-8. 
    A.D.-G. and O.A. acknowledge support from the Spanish Grant No. PID2021-128005NB-C22, funded by MCIN/AEI/10.13039/501100011033 and “ERDF A way of making Europe;” and from Generalitat de Catalunya (No. 2021SGR00856).
    M.S. acknowledges support from Grants No. RYC2022-037932-I and CNS2023-144156 funded by MCIN/AEI/10.13039/501100011033 and the European Union NextGenerationEU/PRTR. M.K. acknowledges support from the ANR funded DATAREDUX project (ANR-19-CE46-0008); the National Laboratory for Health Security, Alfréd Rényi Institute of Mathematics, RRF-2.3.1-21-2022-00006; and the WWTF funded MOMA project (10.47379/ESS22032). J.K. gratefully acknowledges partial support by ERC grant No. 810115-DYNASNET. G.P acknowledges support from the European Research Council (ERC) Consolidator Grant under the European Union’s Horizon Europe programme (grant agreement No. 101171380, project RUNES) and the MSCA Doctoral Network \textit{BeyondTheEdge}(Grant  no. 101120085).
\end{acknowledgements}

\bibliographystyle{unsrt}
\bibliography{bibfile_accelnet}

\end{document}